\documentclass[12pt]{article}
\usepackage{amssymb}
\usepackage{amsmath,bm}
\usepackage{graphics,mathrsfs}
\usepackage{graphicx,epsfig}
\usepackage{subfigure}
\usepackage{placeins}
\usepackage{indentfirst}
\usepackage{setspace}
\usepackage{enumerate}
\usepackage{enumitem}
\usepackage{caption}
\usepackage{varioref}
\usepackage{booktabs,tabularx}
\usepackage{color}
\usepackage[numbers,round]{natbib}
\usepackage{epstopdf}
\setcitestyle{open={[},close={]}}
\usepackage{amsthm}
\makeatletter \oddsidemargin  -.1in \evensidemargin -.1in
\begin{document}
	\newcommand{\bea}{\begin{eqnarray}}
		\newcommand{\eea}{\end{eqnarray}}
	\newcommand{\nn}{\nonumber}
	\newcommand{\bee}{\begin{eqnarray*}}
		\newcommand{\eee}{\end{eqnarray*}}
	\newcommand{\lb}{\label}
	\newcommand{\nii}{\noindent}
	\newcommand{\ii}{\indent}
	\newtheorem{thm}{Theorem}[section]
	\newtheorem{example}{Example}[section]
	\newtheorem{cor}{Corollary}[section]
	\newtheorem{definition}{Definition}[section]
	\newtheorem{lemma}{Lemma}[section]
	\newtheorem{rem}{Remark}[section]
	\newtheorem{proposition}{Proposition}[section]
	\numberwithin{equation}{section}
	\renewcommand{\theequation}{\thesection.\arabic{equation}}
	\renewcommand\bibfont{\fontsize{10}{12}\selectfont}
	\setlength{\bibsep}{0.0pt}
\title{\bf Inference of a competing risks model with partially observed failure causes under improved adaptive type-II progressive censoring* }
\author{ Subhankar {\bf Dutta}\thanks {Email address : subhankar.dta@gmail.com} ~and  Suchandan {\bf  Kayal}\thanks {Email address : (corresponding author)
        kayals@nitrkl.ac.in,~suchandan.kayal@gmail.com,~~~ *It has been published in Proceedings of the Institution of Mechanical Engineers, Part O: Journal of Risk and Reliability, https://doi.org/10.1177/1748006X2211045}
    \\{\it \small Department of Mathematics, National Institute of
        Technology Rourkela, Rourkela-769008, India}}
\date{}
\maketitle
\begin{center}
 \textbf{Abstract}
\end{center}
In this paper, a competing risks model is analyzed based on improved adaptive type-II progressive censored sample (IAT-II PCS). Two independent competing causes of failures are considered.  It is assumed that lifetimes of the competing causes of failure follow  exponential distributions with different means. Maximum likelihood estimators (MLEs) for the unknown model parameters are obtained. Using asymptotic normality property of MLE, the asymptotic confidence intervals are constructed. Existence and uniqueness properties of the MLEs are studied. Further, bootstrap confidence intervals are computed. The Bayes estimators are obtained under symmetric and asymmetric loss functions with non-informative and informative priors. For informative priors, independent gamma distributions are considered.  Highest posterior density (HPD) credible intervals are obtained.  A Monte Carlo simulation study is carried out to compare performance of the established estimates. Furthermore, three different optimality criteria are proposed to obtain the optimal censoring plan. Finally, a real-life data set is considered for illustrative purposes. 

\section{Introduction}
In survival analysis, experimental units may fail due to more than one causes. The experimenter observes time of failure of the experimental units along with the corresponding causes of failure. The causes compete with each other for failure of the experimental units. This kind of studies are known as the competing risks problem in literature. A competing risks data consist of failure time and indicator denoting the cause of failure. There are two mostly used methods to analyze competing risks data. One is latent failure time model proposed by Cox \cite{cox1959analysis} and the other is cause specific hazard rate model proposed by Prentice et. al \cite{prentice1978analysis}. Competing risks data have a wide variety of applications in the fields of engineering, biological and medical studies. For example, a colon cancer patient may die due to other causes than cancer. For various applications of competing risks model, one may refer to Crowder \cite{crowder2012multivariate}.

 Due to improved lifetime of the products, time, and cost constraints, it is often impossible for an experimenter to collect complete sample from a lifetime experiment. As a result, the available data are actually censored. The conventional censoring schemes, namely type-I and type-II have been widely used in reliability studies. Epstein \cite{epstein1954truncated} introduced a mixture of these two censoring schemes, which is known as the type-I hybrid censoring scheme. Childs et. al \cite{childs2003exact} introduced the type-II hybrid censoring scheme. The major drawback of the conventional censoring schemes is that they do not allow to remove experimental units before the terminal point of the experiment. To overcome such drawback,  Cohen \cite{cohen1963progressively} introduced progressive censoring scheme, where one can remove survival units at various stages of an experiment. Extensive developments related to statistical inference of some lifetime distributions based on these censoring schemes have been made in past years. See, for instance, Balakrishnan et. al \cite{balakrishnan2003point},  Balakrishnan and Kundu \cite{balakrishnan2013hybrid}, and Rastogi and Tripathi \cite{rastogi2012estimating}. Note that sometimes the experiment under progressive type-II censoring schemes takes a longer time to get a pre-fixed number of failures. To overcome such drawback, Kundu and Joarder \cite{kundu2006analysis} introduced type-II progressive hybrid censoring scheme. Later, this scheme has been studied by many researchers. See, for example, Lin et. al \cite{lin2009statistical}, Lin et. al \cite{lin2012inference}, Hemmati and Khorram \cite{hemmati2013statistical}, Tomer and Panwar \cite{tomer2015estimation}, and Dutta and Kayal \cite{dutta2022estimation}. Under this censoring scheme, experimental time is prefixed. Here, the number of effective sample size is random. Sometimes, the effective sample size becomes zero and as a result the efficiency of statistical inference decreases. To overcome such inefficiency in experimental results, Ng et. al \cite{ng2009statistical} proposed the adaptive type-II progressive censoring scheme (AT-II PCS). For details about AT-II PCS, one may refer to Ye et. al \cite{ye2014statistical}, Sobhi and Soliman \cite{sobhi2016estimation}, and  El-Sagheer et. al \cite{el2018statistical}.
In AT-II PCS, the experiment will be terminated after getting a prefixed number of failure-time data. Here, the experimental time may be large. To overcome this drawback,  very recently Yan et. al \cite{yan2021statistical} proposed a censoring scheme which is known as the IAT-II PCS. This scheme is described as follows. 

Suppose in an experiment, $n$ number of experimental units are placed. Before starting the experiment, the number of failure, say $m$ and progressive censoring scheme $(R_1,\cdots,R_m)$ are fixed, where $R_m=n-m-\sum_{i=1}^{m-1}R_i$ and $R_i\geq 0$, for $i=1,\cdots,m$. The random lifetime of $i$th failure unit is denoted by $X_{i:m:n}$. In addition, we give two time thresholds $T_1$ and $T_2$ in advance such that $0<T_1<T_2$. Here, the first threshold $T_1$ acts as the warning about the test time. The second theshold $T_2$ represents the maximum duration, to which the experiment will be allowed to continue. It does not matter, whether the experimenter gets $m$ number of failures or not. In this experiment, we mainly have three cases (described below), which have been presented in Figure $1$ for better visualization.  
\begin{align}
	\nonumber \mbox{Case-I}~~~~& X_{m:m:n} < T_{1}<T_{2},\\
	\nonumber \mbox{Case-II}~~~& X_{k_{1}:m:n}<T_{1}<X_{k_{1}+1:m:n}<\cdots<X_{m:m:n}<T_{2},\\
	\nonumber \mbox{Case-III}~~& X_{k_{2}:m:n}<T_{2}<X_{k_{2}+1:m:n}<\cdots<X_{m:m:n}.
\end{align}
 Under the above three cases, we get different sets of failure-times as 
\begin{align}
\nonumber \mbox{Case-I}~~~& \{X_{1:m:n}<\cdots<X_{m:m:n}\},\\
\nonumber \mbox{Case-II}~~& \{X_{1:m:n}<\cdots<X_{k_{1}:m:n}<\cdots<X_{m:m:n}\},\\
\nonumber \mbox{Case-III}~& \{X_{1:m:n}<\cdots<X_{k_{1}:m:n}<\cdots<X_{k_{2}:m:n}\}.
\end{align}
Note that the experiment under IAT-II PCS is terminated at the time $T=min\{X_{m:m:n},T_{2}\}$.
\begin{figure}[h!]
	\begin{center}
		\includegraphics[height=3.6in]{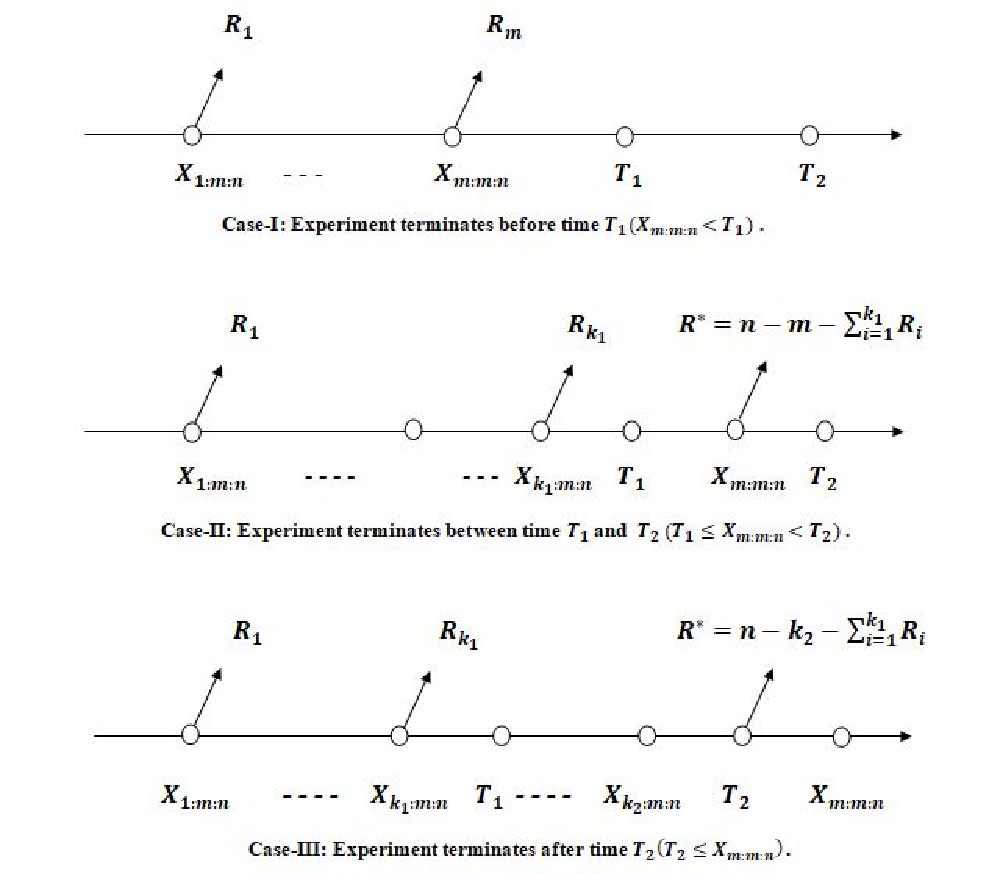}
		\caption{ Schematic representation of the improved adaptive type-II progressive censored scheme. }
	\end{center}
\end{figure}
Various authors have studied competing risks data based on different censoring schemes. Kundu and Basu \cite{kundu2000analysis} analyzed Weibull competing risks in the presence of incomplete data. Kundu and Joarder \cite{kundu2006comp} analyzed the competing risks model based on type-II progressive hybrid censored data. Sarhan \cite{sarhan2007analysis} analyzed competing risks model in the presence of incomplete and censored data based on generalized exponential distribution. Pareek et. al \cite{pareek2009progressively} discussed progressively censored Weibull competing risks model. Cramer \cite{cramer2011progressively} analyzed progressively censored competing risks model based on Lomax distribution. Wang \cite{wang2018inference} discussed progressively censored competing risks model for Kumaraswamy distribution. Ahmed et. al \cite{ahmed2020inference} discussed the inference of progressive type-II competing risks model based on Chen distribution. Hua and Gui \cite{hua2021revisit} considered progressively Type-II censored competing risks model when the latent failure times follow Lomax distributions. Moharib et. al \cite{moharib2021monitoring} monitored the shape parameters of Weibull distributions separately using the control charts based on progressively censored competing risks data. Mahmoud et. al \cite{mahmoud2021inference} considered the estimation of parameters of generalized inverted exponential distribution based on progressively type-I censored competing risks data. Ren and Gui \cite{ren2021inference} discussed the inference based on progressively type-II censored competing risks data where the latent failure times follow generalized Rayleigh distributions. Nassr et. al \cite{nassr2021statistical} discussed the inference based on extended Weibull distributions under adaptive type-II progressively censored competing risks model.    

To the best of our knowledge, till now, no work has been reported based on improved adaptive type-II progressive censored competing risks model in the literature. In this paper, we have considered a very recently proposed censoring scheme (IAT-II PCS) to analyze competing risks data, where the lifetimes of the competing causes of failure follow exponential distributions with different parameters. MLEs are proposed, and their uniqueness and existence properties are studied. Three types of confidence intervals (asymptotic, bootstrap-$p$ and bootstrap-$t$) are constructed. The Bayes estimates are proposed. In this direction, three loss functions are considered. The HPD credible intervals are also obtained. Three different optimality criteria have been proposed to find out the optimal censoring scheme among the chosen schemes. Finally a real-life data set has been analyzed.  

The rest of the paper is organized as follows. In Section $2$, the competing risks model and the corresponding likelihood function based on IAT-II PCS are described.  In Section $3$, the MLEs and their existence and uniqueness properties are discussed. The asymptotic and bootstrap confidence intervals are constructed. The Bayes estimates under different loss functions and the HPD credible intervals are discussed in Section $4$. In Section $5$, a Monte Carlo simulation study is performed to compare the proposed methods. In Section $6$, three different optimality criteria have been proposed. In Section $7$, a real-life data set has been analyzed for illustrative purposes. Finally, conclusion is made in Section $8$.\\

\section{Model description and likelihood function}
In order to analyze the competing risks data, the latent failure time modelling proposed by \cite{cox1959analysis} has been considered in this paper. Two independent competing causes of failures are assumed. Let $X_{i}$ be the lifetime of the $i$th component. Then, 
\begin{align}
\nonumber X_{i}= min\{X_{1i},X_{2i}\},
\end{align}
where $X_{ji}$ $(j=1,2)$ denotes latent failure time of the $i$th experimental unit when $j$th cause of failure occurs. Here, the latent failure times $X_{1i}$ and $X_{2i}$ are independent. Consider ${\delta}_{i}=1$, if the failure occurs due to cause 1 and ${\delta}_{i}=0$, if the failure occurs due to cause 2. Then, $D_{1}=\sum_{i=1}^{D}I({\delta}_{i}=1)$ and  $D_{2}=\sum_{i=1}^{D}I({\delta}_{i}=0)$ denote the total number of failures due to cause 1 and cause 2, respectively, with $D_{1}+D_{2}=D$.
The cumulative distribution function (CDF), probability density function (PDF), survival function, and hazard rate function of latent failure times $X_{ji}$, for $i=1,\cdots,n$ and $j=1,2$ are denoted as $F_{j}(\cdot)$, $f_{j}(\cdot)$, $S_{j}(\cdot)$, and $h_{j}(\cdot),$ respectively. Under this setttings, the likelihood function can be written as
\begin{align}
\nonumber L({\tau}_{1},{\tau}_{2}) \propto \bigg[f_{1}(x_{i}){S_{2}}(x_{i})\bigg]^{D_{1}} \bigg[f_{2}(x_{i}){S_{1}}(x_{i})\bigg]^{D_{2}} \prod_{i=1}^{D}\bigg[{S_{1}}(x_{i}){S_{2}}(x_{i})\bigg]^{R_{i}} \bigg[{S_{1}}(T){S_{2}}(T)\bigg]^{R^{*}},
\end{align}
where $R^*$ is described below. By using $f_{k}(x)=h_{k}(x)S_{k}(x)$, the likelihood function becomes
\begin{align}
L({\tau}_{1},{\tau}_{2}) \propto \bigg[h_{1}(x_{i})\bigg]^{D_{1}} \bigg[h_{2}(x_{i})\bigg]^{D_{2}} \prod_{i=1}^{D}\bigg[{S_{1}}(x_{i}){S_{2}}(x_{i})\bigg]^{1+R_{i}}\bigg[{S_{1}}(T){S_{2}}(T)\bigg]^{R^{*}}, \label{eq 2.1}
\end{align}
where $x_{i}$ represents $i$th failure time $x_{i:m:n}$. Note that for case-I, $D=m$ and $R^{*}= 0$  , for case-II, $D=k_{1}$ and $R^{*}= n-m-\sum_{i=1}^{k_{1}}R_{i}$, and for case-III, $D=k_{1}$ and $R^{*}= n-k_{2}-\sum_{i=1}^{k_{1}}R_{i}$. In this paper, it is assumed that lifetime of the competing causes of failure follows exponential distribution with CDF and hazard rate function, given by
\begin{align}
F_{j}(x)=1-\exp(-{\tau}_{j}x)~~\mbox{and}~~h_{j}(x)={\tau}_{j},\label{eq 2.2}
\end{align}
where $x,{\tau}_{j}>0,$ for $j=1,2$.

\section{Maximum likelihood estimation}
In this section, MLEs of ${\tau}_{1}$ and ${\tau}_{2}$ are obtained. Based on Eqs. $(\ref{eq 2.1})$ and $(\ref{eq 2.2})$, the likelihood function takes the form 
\begin{align}
L({\tau}_{1},{\tau}_{2}) \propto~ {{\tau}_{1}}^{D_{1}} {{\tau}_{2}}^{D_{2}} \prod_{i=1}^{D} \big[\exp\big(-(R_{i}+1)({\tau}_{1}+{\tau}_{2})x_{i}\big)\big] \exp\big(-({\tau}_{1}+{\tau}_{2})T^{*}\big), \label{eq 3.1}
\end{align}
where $T^{*}=TR^{*}$. Now, the log-likelihood function can be expressed as
\begin{align}
\log L({\tau}_{1},{\tau}_{2}) \propto D_{1} \log \tau_{1} +  D_{2} \log \tau_{2} -(\tau_{1}+\tau_{2}) \bigg[\sum_{i=1}^{D} (R_{i}+1)x_{i} +T^{*}\bigg]. \label{eq 3.2}
\end{align}
After differentiating $(\ref{eq 3.2})$ with respect to ${\tau}_{1}$ and ${\tau}_{2}$, and equating them to zero, the likelihood equations are obtained as
\begin{align}
\frac{\partial l}{\partial {\tau}_{1}}=& \frac{D_{1}}{{\tau}_{1}}- \bigg[\sum_{i=1}^{D} (R_{i}+1)x_{i} +T^{*}\bigg]=0, \label{eq 3.3}\\
\frac{\partial l}{\partial {\tau}_{2}}=& \frac{D_{2}}{{\tau}_{2}}- \bigg[\sum_{i=1}^{D} (R_{i}+1)x_{i} +T^{*}\bigg]=0, \label{eq 3.4}
\end{align}
where $l= \log L({\tau}_{1},{\tau}_{2})$. From likelihood equations, the MLEs of $\tau_1$ and $\tau_2$, respectively denoted by  $\widehat{{\tau}_{1}}$ and $\widehat{{\tau}_{2}}$ can be easily obtained, which are given by
\begin{align}
 \widehat{{\tau}_{1}}= \frac{D_{1}}{A}~~~\mbox{and}~~~\widehat{{\tau}_{2}}= \frac{D_{2}}{A}, \label{eq 3.5}
\end{align}
where $A= \sum_{i=1}^{D} (R_{i}+1)x_{i} +T^{*} $. 
\subsection{Existence and uniqueness of MLEs}
In this subsection, we have studied existence and uniqueness of the MLEs. The following theorem presents sufficient conditions such that the MLEs of $\tau_1$ and $\tau_2$ exist uniquely.
\begin{thm}
 For $D_{i}>0$, where $i=1,2$, the MLE for each $\tau_{i}$ exists uniquely.
\end{thm}

\begin{proof}
 From $(\ref{eq 3.5})$, MLEs of the unknown model parameters are
\begin{align}
\nonumber \widehat{{\tau}_{i}}= \frac{D_{i}}{A},~~ i=1,2.
\end{align}
 For the existence of MLE of each model parameter $\tau_{i}$, at least one failure should occur due to $i$th cause, that is, $D_{i}>0$, $i=1,2$. Further, to show uniqueness of the MLEs, we have to establish  that the Hessian matrix $H({\tau}_{1},{\tau}_{2})$ of $l({\tau}_{1},{\tau}_{2})$ is negative definite. Here, the Hessian matrix is
 \begin{align}
 H({\tau}_{1},{\tau}_{2})= \bigg[\frac{\partial^2 l}{\partial {\tau}_{i}\partial {\tau}_{j}} \bigg],~~ i,j=1,2.
 \end{align}
From  $(\ref{eq 3.3})$ and $(\ref{eq 3.4})$, we obtain
\begin{align}
\nonumber \frac{\partial^2 l}{\partial {\tau_{1}}^2}= -\frac{D_{1}}{{\tau_{1}}^2}<0,~~~\frac{\partial^2 l}{\partial {\tau_{2}}^2}= -\frac{D_{2}}{{\tau_{2}}^2}<0,~~\mbox{and}~~ \frac{\partial^2 l}{\partial \tau_{1} \partial \tau_{2}}= 0.
\end{align}
Since the diagonal elements of the Hessian matrix $H({\tau}_{1},{\tau}_{2})$ are negative and rest of the elements are zero, thus Hessian matrix is negative definite, and the result is established.
\end{proof}
\subsection{Asymptotic confidence intervals}
Using asymptotic normality property of the MLEs of ${\tau}_{1}$ and ${\tau}_{2}$, the $100(1-\gamma) \%$ asymptotic confidence intervals are constructed in this subsection. Under some mild regularity conditions, the asymptotic distribution of the MLEs $(\widehat{{\tau}_{1}},\widehat{{\tau}_{2}})^{'}$ can be obtained as
\begin{align}
\nonumber (\widehat{{\tau}_{1}},\widehat{{\tau}_{2}})^{'}-({\tau}_{1},{\tau}_{2})^{'} \sim N\big(0,I^{-1}(\widehat{{\tau}_{1}},\widehat{{\tau}_{2}})\big),
\end{align}
where
\begin{align}\label{3.7}
I^{-1}(\widehat{{\tau}_{1}},\widehat{{\tau}_{2}})= {\begin{bmatrix}
	-l_{20} & -l_{11}\\
	-l_{11} & -l_{02}\\
	\end{bmatrix}}^{-1}_{({\tau}_{1},{\tau}_{2})=(\widehat{{\tau}_{1}},\widehat{{\tau}_{2}})}= {\begin{bmatrix}
	Var(\widehat{{\tau}_{1}}) & Cov(\widehat{{\tau}_{1}},\widehat{{\tau}_{2}})\\
	Cov(\widehat{{\tau}_{1}},\widehat{{\tau}_{2}}) & Var(\widehat{{\tau}_{2}})\\
	\end{bmatrix}},
\end{align}
is the inverse of observed Fisher information matrix for ${\tau}_{1}$ and ${\tau}_{2}$ and $l_{ij}=\frac{\partial^{2} l}{\partial^{i}{\tau}_{1} \partial^{j}{\tau}_{2}}$, $i,j=0,1,2$. Thus, the $100(1-\gamma) \%$ asymptotic confidence intervals for ${\tau}_{1}$ and ${\tau}_{2}$ are given by
\begin{align}
\nonumber \bigg(\widehat{{\tau}_{1}}~ \underline{+}~z_{\frac{\gamma}{2}}\sqrt{Var(\widehat{{\tau}_{1}})}\bigg) ~~\mbox{and}~~ \bigg(\widehat{{\tau}_{2}}~ \underline{+}~z_{\frac{\gamma}{2}}\sqrt{Var(\widehat{{\tau}_{2}})}\bigg),
\end{align}
where $z_{\frac{\gamma}{2}}$ is the upper $\frac{\gamma}{2}$th percentile point of  standard normal distribution.
\subsection{Bootstrap confidence intervals}
Note that to obtain the confidence intervals for small sample size, the bootstrap re-sampling technique provides more accurate result. In this subsection, two different parametric bootstrap confidence intervals have been constructed for unknown parameters. 
\subsubsection{Bootstrap-$p$ (boot-$p$) confidence interval}
Here, the percentile parametric bootstrap confidence intervals have been presented. The following algorithm can be used in this context.\\
------------------------------------------------------------------------------------------------------------------------
\textbf{Algorithm I}\\
------------------------------------------------------------------------------------------------------------------------
\textbf{Step 1:} Generate an IAT-II PCS competing risks sample $(y_1,\cdots,y_D)$ from exponential distribution along with the probabilities of two different assigned causes of failures  $\frac{\widehat{{\tau}_{1}}}{\widehat{{\tau}_{1}}+\widehat{{\tau}_{2}}}$ and $\frac{\widehat{{\tau}_{2}}}{\widehat{{\tau}_{1}}+\widehat{{\tau}_{2}}}$. \\
\textbf{Step 2:} Generate a bootstrap sample using MLEs based on IAT-II PCS and compute the bootstrap MLEs for $\tau_1$ and $\tau_2$, respectively denoted by $\widehat{{\tau}_{1}}^{*}$ and $\widehat{{\tau}_{2}}^{*}$.\\
\textbf{Step 3:} Repeat \textbf{Step 2} for $B$ times to obtain $(\widehat{{\tau}_{1}}^{*(1)},\cdots,\widehat{{\tau}_{1}}^{*(B)})$ and $(\widehat{{\tau}_{2}}^{*(1)},\cdots,\widehat{{\tau}_{2}}^{*(B)})$.\\
\textbf{Step 4:} Rearrange all these MLEs (obtained in \textbf{Step 3}) in ascending order to obtain  $(\widehat{{\tau}_{1}}^{*[1]},\cdots,\widehat{{\tau}_{1}}^{*[B]})$ and $(\widehat{{\tau}_{2}}^{*[1]},\cdots,\widehat{{\tau}_{2}}^{*[B]})$. \\
------------------------------------------------------------------------------------------------------------------------

Now, the $100(1-\gamma)\%$ bootstrap-$p$ confidence intervals for $\tau_{1}$ and $\tau_{2}$ are 
\begin{align}
\nonumber \bigg(\widehat{{\tau}_{1}}^{*[\frac{\gamma B}{2}]},\widehat{{\tau}_{1}}^{*[B-\frac{\gamma B}{2}]}\bigg)~~\mbox{and}~~ \bigg(\widehat{{\tau}_{2}}^{*[\frac{\gamma B}{2}]},\widehat{{\tau}_{2}}^{*[B-\frac{\gamma B}{2}]}\bigg),
\end{align}	 
respectively.
\subsubsection{Bootstrap-$t$ (boot-$t$) confidence interval}
Bootstrap-$t$ confidence intervals are usually constructed when the effective sample size ($m$) is too small. In this regard, one may use the following algorithm.\\
------------------------------------------------------------------------------------------------------------------------
\textbf{Algorithm II}\\
------------------------------------------------------------------------------------------------------------------------
\textbf{Step 1:} Generate an IAT-II PC sample as given in \textbf{Step 1} of bootstrap-$p$ confidence intervals.\\
\textbf{Step 2:} Generate bootstrap samples and the corresponding MLEs $\widehat{{\tau}_{1}}^{*}$ and $\widehat{{\tau}_{2}}^{*}$ as similar  to \textbf{Step 2} in bootstrap-$p$ confidence interval.\\
\textbf{Step 3:} Compute the $t$-statistics for $\tau_{1}$ and $\tau_{2}$, where ${t}_{1}=\frac{\widehat{{\tau}_{1}}^{*}-\widehat{{\tau}_{1}}}{\sqrt{Var(\widehat{{\tau}_{1}}^{*})}}$ and ${t}_{2}=\frac{\widehat{{\tau}_{2}}^{*}-\widehat{{\tau}_{2}}}{\sqrt{Var(\widehat{{\tau}_{2}}^{*})}}$.\\
\textbf{Step 4:} Repeat \textbf{Steps 2-3} for $B$ times and obtain $\big(t_{1}^{(1)},\cdots,t_{1}^{(B)}\big)$ and $\big(t_{2}^{(1)},\cdots,t_{2}^{(B)}\big)$.\\
\textbf{Step 5:} Rearranging the values obtained in \textbf{Step 4} in ascending order, and we obtain $\big(t_{1}^{[1]},\cdots,t_{1}^{[B]}\big)$ and $\big(t_{2}^{[1]},\cdots,t_{2}^{[B]}\big)$.\\
------------------------------------------------------------------------------------------------------------------------
\indent Then, the $100(1-\gamma)\%$ bootstrap-$t$ confidence intervals for $\tau_{1}$ and $\tau_{2}$ are respectively given by 
\begin{align}
\nonumber \bigg(\widehat{{\tau}_{1}}+t_{1}^{[\frac{B\gamma}{2}]},\widehat{{\tau}_{1}}+t_{1}^{[B-\frac{B\gamma}{2}]}\bigg)~~\mbox{and}~~\bigg(\widehat{{\tau}_{2}}+t_{2}^{[\frac{B\gamma}{2}]},\widehat{{\tau}_{2}}+t_{2}^{[B-\frac{B\gamma}{2}]}\bigg).
\end{align}
\section{Bayesian estimation}
In this section, Bayes estimators of the unknown parameters $\tau_{1}$ and $\tau_{2}$ are obtained based on the improved adaptive type-II progressively censored competing risks data. In doing so, three different types of loss functions are considered. As a symmetric loss, the squared error loss function (SELF) is considered. For asymmetric loss, LINEX loss function (LLF), proposed by Varian \cite{varian1975bayesian} and generalized entropy loss function (GELF) are used. Let $\tilde{\theta}$ be an estimator of a parameter $\theta$.  Then, SELF, LLF, and GELF are respectively given by
\begin{eqnarray}
L_{SE}(\tilde{\theta},\theta) &=&(\tilde{\theta}-\theta)^2, \label{eq 4.1}    \\
L_{LI}(\tilde{\theta},\theta)&=&\exp({p(\tilde{\theta}-\theta)})-p(\tilde{\theta}-\theta)-1,~ p\neq 0, \label{eq 4.2} \\
L_{GE}(\tilde{\theta},\theta)&=&\bigg(\frac{\tilde{\theta}}{\theta}\bigg)^q -q\log \bigg(\frac{\tilde{\theta}}{\theta}\bigg)-1,~ q \neq 0. \label{eq 4.3}
\end{eqnarray}
Under the loss functions given by (\ref{eq 4.1}), (\ref{eq 4.2}), and (\ref{eq 4.3}), the Bayes estimates of $\theta$ can be respectively written as the following forms
\begin{eqnarray}
\widehat{\theta}_{SE}&=& E_{\theta}(\theta |\underline{x}), \label{eq 4.4} \\
\widehat{\theta}_{LI}&=& -p^{-1} \log[E_{\theta}(\exp({-p\theta})|\underline{x})],~p\neq 0, \label{eq 4.5}  \\
\widehat{\theta}_{GE}&=& [E_{\theta}({\theta}^{-q}|\underline{x})]^{-\frac{1}{q }},~  q \neq 0,    \label{eq 4.6}
\end{eqnarray}
where the observed data is $\underline{x}=(x_{1},\cdots,x_{D})$.
In the study of Bayesian estimation, choosing priors for the unknown model parameters is an important as well as challenging problem. Here, we consider independent gamma priors as ${\tau}_{1} \sim Gamma(a,b)$ and ${\tau}_{2} \sim Gamma(c,d)$, where $a,b,c,d >0$ are the hyper parameters. It is worth mentioning that  the hyper parameters are chosen to reflect the prior knowledge. Note that the notation $Gamma(c,d)$ implies gamma distribution with shape parameter $c$ and scale parameter $1/d.$ The joint prior probability density function of ${\tau}_{1}$ and ${\tau}_{2}$ can be written as
\begin{eqnarray}
\pi_{1}({\tau}_{1},{\tau}_{2}) \propto {\tau_{1}}^{a-1}  {\tau_{2}}^{c-1} \exp\big[-(b\tau_{1}+d\tau_{2})\big],~\tau_1>0,\tau_2>0. \label{eq 4.7}
\end{eqnarray}
Denote $\Gamma(.)$ as a complete gamma function. Utilizing likelihood function given by $(\ref{eq 3.1})$ and joint prior density function given by $(\ref{eq 4.7})$, the posterior probability density function is obtained as
\begin{eqnarray}
\nonumber\pi({\tau}_{1},{\tau}_{2}|\underline{x})&=& K^{-1}  {{\tau}_{1}}^{a+D_{1}-1} {{\tau}_{2}}^{c+D_{2}-1} \prod_{i=1}^{D} \big[\exp\big(-(R_{i}+1)({\tau}_{1}+{\tau}_{2})x_{i}\big)\big]\\
&~&\times~ \exp\big[-({\tau}_{1}+{\tau}_{2})T^{*}\big] \exp\big[-(b\tau_{1}+d\tau_{2})\big], \label{eq 4.8}
\end{eqnarray}
where
\begin{eqnarray}
\nonumber K&=&\int_{0}^{\infty}\int_{0}^{\infty} {{\tau}_{1}}^{a+D_{1}-1} {{\tau}_{2}}^{c+D_{2}-1} \prod_{i=1}^{D} \big[\exp\big(-(R_{i}+1)({\tau}_{1}+{\tau}_{2})x_{i}\big)\big]\\
\nonumber &~&\times\exp\big[-({\tau}_{1}+{\tau}_{2})T^{*}\big] \exp\big[-(b\tau_{1}+d\tau_{2})\big]~ d\tau_{1} d\tau_{2}\\
\nonumber &=& \frac{\Gamma (a+D_{1})\Gamma (c+D_{1})}{(b+A)^{a+D_{1}}(d+A)^{c+D_{2}}}.
\end{eqnarray}
From $(\ref{eq 4.8})$, the marginal posterior probability density functions of $\tau_{1}$ and $\tau_{2}$ are obtained as
\begin{eqnarray}
\pi(\tau_{1}|\underline{x})&=& \frac{(b+A)^{a+D_{1}}}{\Gamma (a+D_{1})}~{{\tau}_{1}}^{a+D_{1}-1} \exp(-\tau_{1}(b+A)),\label{eq 4.9}\\
\pi(\tau_{2}|\underline{x})&=& \frac{(d+A)^{c+D_{2}}}{\Gamma (c+D_{2})}~{{\tau}_{2}}^{c+D_{2}-1} \exp(-\tau_{2}(d+A)).\label{eq 4.10}
\end{eqnarray}

The following lemma shows that the marginal posterior distributions are log-concave.
\begin{lemma}
 The marginal posterior PDF $\pi(\tau_{k}|\underline{x})$ is log-concave when $D_{k}>0$, for $k=1,2$.
\end{lemma}
\begin{proof}
From $(\ref{eq 4.9})$ and $(\ref{eq 4.10}),$ the marginal posterior PDF can be written as
\begin{eqnarray}
\nonumber \pi(\tau_{k}|\underline{x})\propto {{\tau}_{k}}^{u_{k}+D_{k}-1} \exp(-\tau_{1}(v_{k}+A)),~k=1,2,
\end{eqnarray}
where $(u_{1},u_{2})=(a,c)$ and $(v_{1},v_{2})=(b,d)$.
 Taking logarithm both sides, the above equation reduces to
\begin{eqnarray}
\nonumber \log \pi(\tau_{k}|\underline{x}) \propto (u_{k}+D_{k}-1) \log {\tau}_{k} - {\tau}_{k} (v_{k}+A).
\end{eqnarray}
The first and second order derivatives of  $\log \pi(\tau_{k}|\underline{x})$ with respect to $\tau_k$ are given by
\begin{eqnarray}
\nonumber \frac{\partial \log \pi(\tau_{k}|\underline{x})}{\partial {\tau}_{k}}= \frac{u_{k}+D_{k}-1}{\tau_{k}}-(v_{k}+A)
\end{eqnarray}
and
\begin{eqnarray}
\nonumber \frac{\partial^2 \log \pi(\tau_{k}|\underline{x})}{\partial {{\tau}_{k}}^2}= -\frac{u_{k}+D_{k}-1}{{\tau_{k}}^2} < 0.
\end{eqnarray}
Thus, for each $k=1,2$, $\pi(\tau_{k}|\underline{x})$ is log-concave. 
\end{proof}

Consider an arbitrary function of parameters $\tau_{1}$ and $\tau_{2}$ as $\phi(\tau_{1},\tau_{2})$. Then, from (\ref{eq 4.4}), (\ref{eq 4.5}) and (\ref{eq 4.6}), the Bayes estimates of $\phi(\tau_{1},\tau_{2})$ with respect to SELF, LLF and GELF are given by
\begin{eqnarray}
\widehat{\phi}_{SE}&=&~ {K}^{-1}\int_{0}^{\infty}\int_{0}^{\infty} \phi(\tau_{1},\tau_{2}) \pi(\tau_{1},\tau_{2} |\underline{x})~ d\tau_{1} d\tau_{2}, \label{eq 4.11}\\
\widehat{\phi}_{LI}&=&~-\bigg(\frac{1}{p}\bigg)\log \bigg[{K}^{-1} \int_{0}^{\infty}\int_{0}^{\infty}e^{-p\phi(\tau_{1},\tau_{2})} \pi(\tau_{1},\tau_{2} |\underline{x})~ d\tau_{1} d\tau_{2}\bigg], \label{eq 4.12} \\
\widehat{\phi}_{GE}&=&~ \bigg[{K}^{-1}\int_{0}^{\infty}\int_{0}^{\infty} (\phi(\tau_{1},\tau_{2}))^{-q} \pi(\tau_{1},\tau_{2} |\underline{x})~ d\tau_{1} d\tau_{2}\bigg]^{-\frac{1}{q}},  \label{eq 4.13}
\end{eqnarray}
respectively. Replacing $\phi(\tau_{1},\tau_{2})$ as $\tau_{1}$ and $\tau_{2}$ in $(\ref{eq 4.11})$, $(\ref{eq 4.12})$, and $(\ref{eq 4.13}),$ the Bayes estimates of $\tau_{1}$ and $\tau_{2}$ can be obtained.
Using $(\ref{eq 4.11})$, $(\ref{eq 4.12})$ and $(\ref{eq 4.13})$, the Bayes estimates of $\tau_1$ and $\tau_2$ under SELF can be respectively obtained as
\begin{eqnarray}
 \widehat{{\tau}_{1}}_{SE}=~ \frac{a+D_{1}}{b+A}~~~\mbox{and}~~~\widehat{{\tau}_{2}}_{SE}=~ \frac{c+D_{2}}{d+A}.
\end{eqnarray}
The Bayes estimates of $\tau_1$ and $\tau_2$ under LLF are respectively given by
\begin{align}
 \widehat{{\tau}_{1}}_{LI}=-\frac{a+D_{1}}{p}~\log \bigg(\frac{b+A}{p+b+A}\bigg)~~\mbox{and}~~ \widehat{{\tau}_{2}}_{LI}=-\frac{c+D_{2}}{p}~\log \bigg(\frac{d+A}{p+d+A}\bigg).
\end{align}
Further, the Bayes estimates of $\tau_1$ and $\tau_2$ under GELF can be obtained, respectively as
\begin{eqnarray}
 \widehat{{\tau}_{1}}_{GE}=\frac{1}{b+A} \bigg(\frac{\Gamma(a+D_{1}-q)}{\Gamma(a+D_{1})}\bigg)~~\mbox{and}~~\widehat{{\tau}_{2}}_{GE}=\frac{1}{d+A} \bigg(\frac{\Gamma(c+D_{2}-q)}{\Gamma(c+D_{2})}\bigg).
\end{eqnarray}

Note that the marginal posterior distribution follows gamma distribution which is unimodal. Since the posterior distribution is unimodal, the $100(1-\gamma)\%$ HPD credible intervals can be obtained by solving the following equations 
\begin{align}
	\nonumber \int_{l_{k}}^{u_{k}}\pi(\tau_{k}|data) = 1- \gamma ~~~
\mbox{and} ~~~\pi(l_{k}|data)=   \pi(u_{k}|data),
\end{align}  
where $l_{k}$ and $u_{k}$, for $k=1,2$ are the lower and upper bounds of the HPD credible intervals.

\section{Simulation study}
In this section, a Monte Carlo simulation study has been employed to compare the performance of the proposed estimates based on 10,000 generated IAT-II PCS using $R$ software. The performance of  proposed point estimates are compared based on  mean squared error (MSE). The performance of the proposed interval estimates are compared with respect to average length (AL) and coverage probability (CP).
\begin{itemize}
	\item \textbf{MSE:} It is given by $\frac{1}{N}\sum_{i=1}^{N}(\widehat{{\tau}_{i}}-\tau_{i})^2$, for $i=1,2$, where $N$ denotes the number of replications in the Monte carlo simulation. The smaller value of MSE represents an estimator with better accuracy.
	\item \textbf{AL:} The smaller length of the interval estimates indicates that the prediction model represents better accuracy with the experimental data.
	\item \textbf{CP:} The probability that true parameter value is contained in the interval estimates. When the value of CP is nearly about the nominal value, that is, $100(1-\gamma)\%$, then it provides better result in terms of CP.
\end{itemize}
In order to generate samples based on IAT-II PCS with competing risks from exponential distribution, the following algorithm is proposed.\\
------------------------------------------------------------------------------------------------------------------------
\textbf{Algorithm III}\\
------------------------------------------------------------------------------------------------------------------------
\textbf{Step 1:} Generate progressive type-II censored exponential data of sample size $m$ with parameter $\tau_{1}+\tau_{2}$. \\
\textbf{Step 2:} Obtain IAT-II PCS of size $D$ by using $X_{m:m:n}$ two time thresholds $T_1$ and $T_2$.\\
\textbf{Step 3:} Assign failure risk $\delta_{i}=0,1,(i=1,\cdots,m)$, to the generated censored data.The probability of failure risks are assigned as $P(\delta_{i}=1)=\frac{\tau_{1}}{\tau_{1}+\tau_{2}}$ and $P(\delta_{i}=0)=\frac{\tau_{2}}{\tau_{1}+\tau_{2}}$.\\
------------------------------------------------------------------------------------------------------------------------ 

 For the purpose of simulation, different sample sizes $(n)$, different effective sample sizes $(m)$, different time thresholds $(T_{1},T_{2}),$ and different censoring schemes have been considered. The censoring schemes (CS) are\\[2 mm]
\textbf{Scheme-I}~~~$R_{1}=R_{2}=\cdots=R_{m-1}=0,R_{m}=n-m$.\\
\textbf{Scheme-II}~~$R_{1}=R_{2}=\cdots=R_{m-1}=1,R_{m}=n-2m+1$.\\
\textbf{Scheme-III}~$R_{1}=R_{2}=\cdots=R_{m}=(n-m)/m$.\\[1 mm]
The average values and MSEs of the MLEs and Bayes estimates are computed when the true values of $(\tau_{1},\tau_{2})$ are taken as $(0.6,0.8)$ and $(1,1.5)$. Note that in the simulation study the Bayes estimates are calcuated based on non-informative prior (NIP) and informative prior (IP).  In case of NIP, the hyper parameters are assumed as zero and for IP, the hyper parameters are obtained by using the following formula (also, see Dey et. al \cite{dey2016estimation}) 
 \begin{eqnarray*}
	a=\frac{\widehat{\tau_{1}}^2}{\frac{1}{N-1}\sum_{i=1}^{N}(\widehat{\tau}_{1i}-\widehat{\tau_{1}})^2},~~b=\frac{\widehat{\tau_{1}}}{\frac{1}{N-1}\sum_{i=1}^{N}(\widehat{\tau}_{1i}-\widehat{\tau_{1}})^2},\\c=\frac{\widehat{\tau_{2}}^2}{\frac{1}{N-1}\sum_{i=1}^{N}(\widehat{\tau}_{2i}-\widehat{\tau_{2}})^2},~~d=\frac{\widehat{\tau_{2}}}{\frac{1}{N-1}\sum_{i=1}^{N}(\widehat{\tau}_{2i}-\widehat{\tau_{2}})^2},
\end{eqnarray*}
 where $\widehat{\tau}_{1i}$ and $\widehat{\tau}_{2i}$, $i=1,\cdots,N$, are the MLEs of $\tau_{1}$ and $\tau_{2}$, respectively for $N$ number of IAT-II PCSs. Bayes estimates are obtained under SELF, LLF (when $p=-0.05, 0.5$), and GELF (when $q=-0.05, 0.5$). Further, confidence intervals and HPD credible intervals are compared on the basis of AL and CP. From Tables $\ref{T1},~\ref{T3},~\ref{T5}$, and $\ref{T7}$, some observations are made, which are presented below.
\begin{itemize}
	\item For fixed value of $n$, when effective sample size $m$ increases, the simulated MSEs decrease. For clear visualisation, for Table $1,$ we have plotted the MSE values of different estimates in Figure $2$. In Figure $2$, along horizontal axis, total eleven estimates as presented in Table $1$ are considered. For example, the bar plots corresponding to $1$ represents the MSEs of MLE, corresponding to $2$ represents the MSEs of Bayes estimates under SELF with respect to NIP, and so on.
	\item  Bayes estimates based on NIP perform better than MLEs in terms of MSEs. However, the Bayes estimates based on IP perform better than that with respect to NIP.
	\item Based on average values and MSEs, the Bayes estimates under LLF  and GELF provide better results than other estimates for $\tau_{1}$ and $\tau_{2}$, respectively.
	\item When the values of time thresholds $T_{1}$ and $T_{2}$ increase, then MSEs decrease.
\end{itemize}
In case of interval estimates, from Tables $\ref{T2},~\ref{T4},~\ref{T6}$, and $\ref{T8},$ the following observations are pointed out.
\begin{itemize}
	\item When sample size ($n$) and effective sample size ($m$) increase, ALs of intervals decrease.
	\item Based on AL and CP, the HPD credible intervals perform better than asymptotic and bootstrap confidence intervals. Boot-$p$ confidence intervals perform better than boot-$t$ confidence intervals and asymptotic confidence intervals (ACI) in terms of AL.
	\item In terms of CP, boot-$t$ confidence intervals perform better than other confidence and HPD credible intervals.
	\item When the values of time thresholds $T_{1}$ and $T_{2}$ increase, the length of intervals decrease.
\end{itemize}
The performance of classical and Bayesian estimates are quite satisfactory. The values of point estimates under MLE and Bayes estimates with respect to NIP under SELF are same upto the reported (here three) decimal places. Further, when the value of $p$ is nearly zero, the Bayes estimates under LLF are almost equal to the values of Bayes estimates under SELF. The Bayes estimates with respect to IP under LLF and GELF are superior choices in the present problem. It has been observed that censoring Scheme-I provides the smallest MSE and AL among three censoring schemes. Further, HPD credible intervals are recommended to use if one considers AL. However, if we consider CP, the boot-$t$ confidence intervals are recommended to use.

\begin{table}[htbp!]
	\begin{center}
		\caption{Average values and MSEs of MLEs and Bayes estimates for different schemes with different values of $n$, $m$ when $T_{1}=0.5$, $T_{2}=1$, and $\Theta=(\tau_{1},\tau_{2})=(0.6,0.8)$.}
		\label{T1}
		\tabcolsep 7pt
		\small
		\scalebox{0.74}{
			\begin{tabular}{*{14}c*{13}{r@{}l}}
				\toprule
				\multicolumn{4}{c}{} &
				\multicolumn{2}{c}{SELF} & \multicolumn{2}{c}{LLF ($p=-0.05$)} & \multicolumn{2}{c}{LLF ($p=0.5$)}  &
				\multicolumn{2}{c}{GELF ($q=-0.05$)} & \multicolumn{2}{c}{GELF ($q=0.5$)}  &\\
				\cmidrule(lr){5-6}\cmidrule(lr){7-8} \cmidrule(lr){9-10} \cmidrule(lr){11-12} \cmidrule(lr){13-14}
				\multicolumn{1}{c}{$(n,m)$}&\multicolumn{1}{c}{CS} & \multicolumn{1}{c}{$\Theta$}& \multicolumn{1}{c}{MLE} & \multicolumn{1}{c}{NIP}
				& \multicolumn{1}{c}{IP} & \multicolumn{1}{c}{NIP}
				& \multicolumn{1}{c}{IP} & \multicolumn{1}{c}{NIP}
				& \multicolumn{1}{c}{IP} & \multicolumn{1}{c}{NIP}
				& \multicolumn{1}{c}{IP} &\multicolumn{1}{c}{NIP}
				& \multicolumn{1}{c}{IP} \\
				\midrule
				(30,10)& I& $\tau_{1}$& 0.622& 0.622& 0.596& 0.625& 0.598&  0.597& 0.583& 0.550&  0.556& 0.506& 0.532 \\
				& & & 0.051& 0.051& 0.012& 0.052& 0.012& 0.041& 0.011& 0.042& 0.012& 0.042& 0.014 \\
				& & $\tau_{2}$& 0.935& 0.935& 0.853& 0.939& 0.855& 0.896& 0.835& 0.862& 0.812& 0.819& 0.789 \\
				& & & 0.129& 0.129& 0.027& 0.133& 0.027& 0.101& 0.023& 0.099&  0.022& 0.086& 0.020\\[2 mm]			
				& II& $\tau_{1}$& 0.621& 0.621& 0.598& 0.632&  0.599& 0.602& 0.585& 0.555& 0.558& 0.511& 0.534\\
				& & & 0.059& 0.059& 0.014&  0.061& 0.015& 0.049& 0.014& 0.049& 0.015& 0.048& 0.016 \\
				& & $\tau_{2}$& 0.958& 0.958& 0.863& 0.963& 0.865& 0.918& 0.845& 0.883& 0.823& 0.839& 0.798\\
				& & & 0.146& 0.146& 0.030& 0.150& 0.030& 0.114& 0.025& 0.110& 0.024& 0.095& 0.021\\[2 mm]				
				& III& $\tau_{1}$& 0.538& 0.538& 0.548& 0.541& 0.549& 0.515& 0.536& 0.463& 0.508& 0.418&  0.483\\
				& & & 0.057& 0.057& 0.020& 0.058& 0.020& 0.053& 0.020& 0.061& 0.024&  0.071& 0.028\\
				& & $\tau_{2}$& 0.922& 0.922& 0.839& 0.926& 0.842& 0.882& 0.821& 0.846&  0.798& 0.801& 0.774\\
				& & & 0.149& 0.149& 0.039& 0.152& 0.039& 0.122& 0.035& 0.119& 0.034& 0.107& 0.033 \\[2 mm]				
		\midrule
		(30,15)& I& $\tau_{1}$& 0.587& 0.587& 0.583& 0.589& 0.584& 0.572& 0.574& 0.539& 0.552& 0.510& 0.533\\
		& & & 0.032& 0.032& 0.012& 0.032&  0.012& 0.029& 0.011& 0.032& 0.011& 0.033& 0.013 \\
		& & $\tau_{2}$& 0.876& 0.876& 0.838& 0.879& 0.840& 0.853& 0.824& 0.828& 0.807&  0.799& 0.788\\
		& & &  0.076&  0.076& 0.025& 0.077& 0.025& 0.066& 0.020& 0.065& 0.020& 0.061& 0.019\\[2 mm]		
		& II& $\tau_{1}$& 0.529& 0.529& 0.541& 0.531& 0.542& 0.513& 0.531& 0.474&  0.507& 0.442&  0.487\\
		& & & 0.057& 0.057& 0.013& 0.058& 0.013& 0.055& 0.013& 0.047& 0.014& 0.047& 0.015\\
		& & $\tau_{2}$& 0.821& 0.821& 0.794& 0.824& 0.796& 0.796& 0.780& 0.767& 0.761& 0.734& 0.741\\
		& & & 0.097& 0.097& 0.025& 0.098& 0.025& 0.087& 0.023& 0.089& 0.022& 0.087& 0.020 \\[2 mm]		
		& III & $\tau_{1}$& 0.522& 0.522& 0.539& 0.524& 0.539& 0.506& 0.528& 0.468& 0.504& 0.436& 0.484\\
		& & & 0.050& 0.050&  0.020& 0.050& 0.020& 0.048& 0.020 & 0.055& 0.024& 0.062& 0.028\\
		& & $\tau_{2}$& 0.809& 0.809& 0.789& 0.812& 0.791& 0.784& 0.774& 0.755& 0.756& 0.723& 0.735  \\
		& & & 0.085& 0.085& 0.031& 0.086& 0.031& 0.076& 0.029& 0.079& 0.031& 0.078& 0.032 \\[2 mm]
		\midrule
		(40,10)& I& $\tau_{1}$&0.620& 0.620& 0.590& 0.622& 0.592& 0.594& 0.583& 0.548& 0.555& 0.505& 0.533 \\
		& & & 0.055& 0.055& 0.011& 0.057& 0.011& 0.045& 0.010& 0.045& 0.011& 0.040& 0.012 \\
		& & $\tau_{2}$& 0.929& 0.929& 0.847& 0.934& 0.849& 0.891& 0.829& 0.857& 0.808& 0.814& 0.784\\
		& & & 0.140& 0.140& 0.027& 0.144& 0.028& 0.109& 0.024& 0.109& 0.023& 0.095& 0.021 \\[2 mm]		
		& II& $\tau_{1}$& 0.615& 0.615& 0.592& 0.617& 0.593& 0.590& 0.579& 0.543& 0.552& 0.501& 0.529 \\
		& & & 0.045& 0.045& 0.012 & 0.046& 0.012& 0.038& 0.011& 0.039& 0.012& 0.039& 0.014 \\
		& & $\tau_{2}$& 0.923& 0.923& 0.846& 0.927& 0.848& 0.885& 0.828& 0.851& 0.806& 0.808& 0.783 \\
		& & & 0.117& 0.117&0.026 & 0.120& 0.026& 0.092& 0.023& 0.089& 0.022& 0.078& 0.021 \\[2 mm]		
		 & III& $\tau_{1}$& 0.580& 0.580& 0.570& 0.579& 0.570 & 0.552& 0.556 & 0.503& 0.529 & 0.458&0.505  \\
		 & & & 0.055& 0.055& 0.018& 0.056& 0.018& 0.051& 0.017& 0.056& 0.020& 0.061& 0.023  \\
		 & & $\tau_{2}$& 0.948& 0.948& 0.852 & 0.943&0.850 & 0.907&0.834 & 0.865&0.807 & 0.828& 0.788 \\
		 & & & 0.143& 0.143& 0.032& 0.140& 0.031& 0.113& 0.028& 0.106& 0.026& 0.095& 0.025  \\[2 mm]
		 \midrule		
		(40,20)& I& $\tau_{1}$&0.581& 0.581& 0.580 & 0.580& 0.579 &0.570& 0.572 & 0.542& 0.551& 0.524& 0.539 \\
        & & & 0.024& 0.024& 0.011 & 0.023 & 0.011 & 0.022& 0.011 & 0.024&0.012 & 0.025& 0.013  \\	
        & & $\tau_{2}$& 0.868& 0.868& 0.840 & 0.866& 0.838& 0.851& 0.828& 0.828& 0.811 & 0.811& 0.799    \\	
        & & & 0.057& 0.057& 0.026& 0.056& 0.025& 0.051& 0.024& 0.049& 0.023& 0.047& 0.021 \\[2 mm]
        & II& $\tau_{1}$& 0.536& 0.536& 0.544 & 0.535& 0.543 & 0.524& 0.536 & 0.491& 0.513 & 0.471& 0.499\\
        & & & 0.036& 0.036& 0.011& 0.035& 0.011& 0.034& 0.011& 0.039& 0.012& 0.038& 0.012  \\
        & & $\tau_{2}$& 0.816& 0.816& 0.791& 0.809& 0.793& 0.792& 0.785& 0.771& 0.769& 0.749& 0.748 \\
        & & & 0.071& 0.071& 0.023& 0.070& 0.023& 0.066& 0.021& 0.067& 0.021& 0.067& 0.020  \\[2 mm]
        & III& $\tau_{1}$& 0.538& 0.538& 0.557& 0.538& 0.551& 0.535& 0.542& 0.492& 0.518& 0.471& 0.508  \\
        & & & 0.035& 0.035& 0.016& 0.036& 0.016& 0.033& 0.015& 0.038& 0.018& 0.042& 0.021\\
        & & $\tau_{2}$& 0.832& 0.832& 0.798& 0.805& 0.792& 0.789& 0.787& 0.766& 0.768& 0.743& 0.751   \\
        & & & 0.070& 0.070& 0.028& 0.069& 0.028& 0.061& 0.027& 0.067& 0.025& 0.066& 0.024 \\[2 mm]
        \midrule				
\end{tabular}}
\end{center}
\vspace{-0.5cm}
\end{table}

\begin{table}[htbp!]
	\begin{center}
		\caption{ AL and CP of $95\%$ confidence and HPD credible intervals of $\tau_{1}$, and $\tau_{2}$ for different schemes with different values of $n$, $m$ when $T_{1}=0.5$, $T_{2}=1$, and $\Theta=(\tau_{1},\tau_{2})=(0.6,0.8)$. }
		\label{T2}
		\tabcolsep 7pt
		\small	
			\begin{tabular}{*{11}c*{10}{r@{}l}}
				\toprule
				\multicolumn{3}{c}{} &
				\multicolumn{2}{c}{ACI} & \multicolumn{2}{c}{{Boot-$p$}} & \multicolumn{2}{c}{{Boot-$t$}} & \multicolumn{2}{c}{HPD} & \\
				\cmidrule(lr){4-5}\cmidrule(lr){6-7} \cmidrule(lr){8-9} \cmidrule(lr){10-11}
				\multicolumn{1}{c}{$(n,m)$}& \multicolumn{1}{c}{CS} & \multicolumn{1}{c}{$\Theta$}& \multicolumn{1}{c}{AL} & \multicolumn{1}{c}{CP}
				& \multicolumn{1}{c}{AL} & \multicolumn{1}{c}{CP}
				& \multicolumn{1}{c}{AL} & \multicolumn{1}{c}{CP} &  \multicolumn{1}{c}{AL} & \multicolumn{1}{c}{CP} \\		
				\midrule
				(30,10)& I& $\tau_{1}$ &1.214& 0.975& {0.801}& {0.978}& {0.834}& {0.975}& {0.413}& {0.987}\\[1 mm]
				& & $\tau_{2}$& 1.489& 0.993& {1.289}& {0.974}& {1.311}& {0.969}& {0.686}& {0.985} \\[2 mm]	
				& II& $\tau_{1}$& 1.235& 0.969& {0.836}& {0.970}& {0.887}& {0.962}& {0.480} & {0.982} \\
				& & $\tau_{2}$& 1.523& 0.986& {1.234}& {0.972}&
				{1.318}& {0.969}& {0.666}& {0.987}\\[2 mm]	
				& III& $\tau_{1}$& 1.165& 0.902& {0.810}& {0.961}&
				{0.875}& {0.956}& {0.564}& {0.967}\\[1 mm]
				& & $\tau_{2}$& 1.519& 0.894& {1.402}& {0.962}&
				{1.421}& {0.960}& {0.752}& {0.970}\\[2 mm]	
			\midrule
			(30,15)& I& $\tau_{1}$ & 0.956& 0.962& {0.692}& {0.962}&
			{0.714}& {0.955}& {0.447}& {0.976} \\[1 mm]
			& & $\tau_{2}$& 1.172& 0.976& {0.955}& {0.971}&
			{1.010}& {0.968}& {0.623}& {0.986}\\[2 mm]
			& II& $\tau_{1}$ & 0.994& 0.890& {0.736}& {0.958}& {0.788}& {0.952}& {0.498}& {0.968} \\[1 mm]
			& & $\tau_{2}$& 1.238& 0.941& {1.086}& {0.958}&
			{1.124}& {0.956}& {0.724}& {0.966}\\[2 mm]
			& III& $\tau_{1}$& 0.971& 0.870& {0.739}& {0.962}&
			{0.801}& {0.958}& {0.507}& {0.974} \\[1 mm]
			& & $\tau_{2}$& 1.213& 0.936& {0.917}& {0.959}&
			{0.933}& {0.955}& {0.699}& {0.969}\\[2 mm]
			\midrule
			(40,10)& I& $\tau_{1}$ & 1.223& 0.986& {0.793}& {0.968}&
			{0.814}& {0.961}& {0.418}& {0.979}\\[1 mm]
			& & $\tau_{2}$& 1.499& 0.987& {1.179}& {0.966}&
			{1.210}& {0.963}& {0.597}& {0.972} \\[2 mm]
			& II& $\tau_{1}$ & 1.229& 0.992& {0.791}& {0.952}&
			{0.834}& {0.949}& {0.415}& {0.968} \\[1 mm]
			& & $\tau_{2}$& 1.506& 0.993& {1.078}& {0.967}&
			{1.153}& {0.960}& {0.592}& {0.977} \\[2 mm]
			& III& $\tau_{1}$ & 1.194& 0.942& {0.836}& {0.961}&
			{0.891}& {0.954}& {0.546}& {0.973} \\[1 mm]
			& & $\tau_{2}$& 1.530&  0.973& {1.214}& {0.960}&
			{1.287}& {0.952}& {0.737}& {0.975}\\[2 mm]
			\midrule	
			(40,20)& I& $\tau_{1}$ & 0.815& 0.968& {0.585}& {0.955}&
			{0.634}& {0.948}& {0.392}& {0.966}\\[1 mm]
			& & $\tau_{2}$ & 0.997& 0.972& {0.882}& {0.961}&
			{0.911}& {0.956}& {0.584}& {0.969}\\[2 mm]
			& II& $\tau_{1}$ & 0.848& 0.921& {0.639}& {0.941}&
			{0.660}& {0.945}& {0.504}& {0.968} \\[1 mm]
			& & $\tau_{2}$ & 1.040& 0.924& {0.922}& {0.954}&
			{0.949}& {0.948}& {0.715}& {0.963}\\[2 mm]
			& III& $\tau_{1}$ & 0.835&  0.935& {0.645}& {0.959}&
			{0.690}& {0.953}& {0.507}& 0.977\\[1 mm]
			& & $\tau_{2}$ & 1.022&  0.940& 0.932& 0.965&
			0.973& 0.955& 0.706& 0.975 \\[2 mm]
			\midrule
		\end{tabular}
	\end{center}
	\vspace{-0.5cm}
\end{table}

\begin{table}[htbp!]
	\begin{center}
		\caption{Average values and MSEs of MLEs and Bayes estimates for different schemes with different values of $n$, $m$ when $T_{1}=0.5$, $T_{2}=1$, and $\Theta=(\tau_{1},\tau_{2})=(1,1.5)$. }
		\label{T3}
		\tabcolsep 7pt
		\small
		\scalebox{0.74}{
			\begin{tabular}{*{14}c*{13}{r@{}l}}
				\toprule
				\multicolumn{4}{c}{} &
				\multicolumn{2}{c}{SELF} & \multicolumn{2}{c}{LLF ($p=-0.05$)} & \multicolumn{2}{c}{LLF ($p=0.5$)}  &
				\multicolumn{2}{c}{GELF ($q=-0.05$)} & \multicolumn{2}{c}{GELF ($q=0.5$)}  &\\
				\cmidrule(lr){5-6}\cmidrule(lr){7-8} \cmidrule(lr){9-10} \cmidrule(lr){11-12} \cmidrule(lr){13-14}
				\multicolumn{1}{c}{$(n,m)$}&\multicolumn{1}{c}{CS} & \multicolumn{1}{c}{$\Theta$}& \multicolumn{1}{c}{MLE} & \multicolumn{1}{c}{NIP}
				& \multicolumn{1}{c}{IP} & \multicolumn{1}{c}{NIP}
				& \multicolumn{1}{c}{IP} & \multicolumn{1}{c}{NIP}
				& \multicolumn{1}{c}{IP} & \multicolumn{1}{c}{NIP}
				& \multicolumn{1}{c}{IP} &\multicolumn{1}{c}{NIP}
				& \multicolumn{1}{c}{IP} \\
				\midrule
			   (30, 10)& I& $\tau_{1}$& 1.102& 1.102& 1.110& 1.111& 1.108 & 1.025& 1.003 & 0.974& 0.969 & 0.898& 0.951\\
			   & & & 0.163& 0.163& 0.069 & 0.170& 0.067 & 0.110& 0.066& 0.119& 0.067& 0.111& 0.068   \\
			   & & $\tau_{2}$& 1.653& 1.653& 1.570& 1.666& 1.561& 1.538& 1.511& 1.524& 1.480& 1.448& 1.452 \\
			   & & & 0.366& 0.366& 0.103 & 0.383&0.102 & 0.249& 0.087& 0.292&0.091 & 0.265&0.090  \\[2 mm]	
			   & II & $\tau_{1}$& 1.109& 1.109&  1.086&  1.118& 1.014&  1.031& 1.007 & 0.980& 0.965 & 0.903& 0.923\\
			   & & & 0.178& 0.178& 0.068& 0.187& 0.067& 0.120& 0.064& 0.131& 0.064& 0.120& 0.066  \\
			   & & $\tau_{2}$& 1.663& 1.663& 1.562& 1.676& 1.578& 1.546& 1.535& 1.534& 1.489& 1.457& 1.493  \\
			   & & & 0.402& 0.402& 0.171& 0.421& 0.170 & 0.271& 0.162& 0.320& 0.158& 0.290& 0.155 \\[2 mm]
			    & III & $\tau_{1}$& 1.100& 1.100& 1.042& 1.109& 1.041& 1.020& 0.997& 0.967& 0.965& 0.888& 0.918 \\
			    & & & 0.193& 0.193& 0.058& 0.201& 0.057& 0.135& 0.052& 0.148& 0.053& 0.139& 0.054 \\
			    & & $\tau_{2}$& 1.714& 1.714& 1.611& 1.728& 1.602& 1.591& 1.551& 1.580& 1.513& 1.506& 1.493\\
			    & & & 0.399& 0.399& 0.112& 0.419& 0.116& 0.263& 0.091& 0.307& 0.092& 0.272& 0.090 \\
			   \midrule
			   (30, 15)& I& $\tau_{1}$& 1.065& 1.065& 1.015& 1.070& 1.021& 1.018& 1.011& 0.982& 0.986& 0.933& 0.948 \\
			   & & &  0.084&  0.084& 0.040& 0.086& 0.041& 0.066& 0.037& 0.068& 0.036& 0.066& 0.037   \\
			   & & $\tau_{2}$& 1.598& 1.598& 1.525& 1.606& 1.517& 1.527& 1.498& 1.514& 1.487& 1.465& 1.474   \\
			   & & & 0.189& 0.189& 0.092 & 0.194& 0.090& 0.148& 0.083& 0.161& 0.083& 0.152& 0.082  \\[2 mm]
			   & II & $\tau_{1}$& 1.005& 1.005& 1.011& 1.011& 0.989& 0.956& 0.964& 0.915&  0.933& 0.862& 0.894   \\
			   & & & 0.123& 0.123& 0.060& 0.126& 0.061& 0.106& 0.059& 0.114& 0.060& 0.117& 0.061 \\
			   & & $\tau_{2}$& 1.669& 1.669& 1.591& 1.677& 1.597& 1.587& 1.527& 1.578& 1.525& 1.525& 1.491  \\
			   & & & 0.301& 0.301& 0.141& 0.309& 0.143& 0.236& 0.122& 0.252& 0.126& 0.232& 0.125   \\[2 mm]
			   & III & $\tau_{1}$& 1.012& 1.012& 0.989& 1.017& 0.992& 0.962& 0.956& 0.921& 0.925& 0.868& 0.885  \\
			   & & & 0.116& 0.116& 0.049& 0.118& 0.051& 0.099& 0.046& 0.106& 0.047& 0.109& 0.048 \\
			   & &  $\tau_{2}$& 1.679& 1.679& 1.588& 1.688& 1.605& 1.597& 1.545& 1.588& 1.530& 1.534& 1.497 \\
			   & & & 0.288& 0.288& 0.097& 0.296& 0.102& 0.224& 0.078& 0.239& 0.081&  0.219& 0.078 \\[2 mm]
			   \midrule
			  (40, 10)& I& $\tau_{1}$& 1.118& 1.118& 1.055& 1.127& 1.062& 1.039& 1.010& 0.989& 0.961& 0.911& 0.919 \\
			  & & & 0.170& 0.170& 0.050& 0.178& 0.052& 0.115& 0.040& 0.122& 0.041& 0.111& 0.042 \\
			  & & $\tau_{2}$& 1.678& 1.678& 1.571& 1.691& 1.572& 1.559& 1.509& 1.547& 1.482& 1.470& 1.449 \\
			  & & & 0.382& 0.382& 0.108& 0.400& 0.112& 0.259& 0.088& 0.300& 0.090& 0.270& 0.089 \\[2 mm]
			  & II& $\tau_{1}$& 1.101& 1.101& 1.023& 1.110& 1.040& 1.025& 0.989& 0.973& 0.950& 0.897& 0.903   \\
			  & & & 0.156& 0.156& 0.046& 0.163& 0.048& 0.107& 0.041& 0.115& 0.043& 0.107& 0.044 \\
			  & & $\tau_{2}$& 1.652& 1.652& 1.529& 1.665& 1.571& 1.537& 1.508& 1.523& 1.479& 1.447& 1.451   \\
			  & & & 0.351& 0.351& 0.109& 0.367& 0.112& 0.241& 0.095& 0.279& 0.094& 0.254& 0.093 \\[2 mm]
			  & III& $\tau_{1}$& 1.126& 1.126& 1.057& 1.135& 1.068& 1.044& 1.007& 0.994& 0.979& 0.915& 0.926 \\
			  & & & 0.204& 0.204& 0.058& 0.213& 0.060& 0.137& 0.048& 0.148& 0.049& 0.133& 0.050  \\
			  & & $\tau_{2}$& 1.705& 1.705& 1.593& 1.719& 1.604& 1.582& 1.511& 1.572& 1.504& 1.494& 1.489   \\
			  & & & 0.445& 0.445& 0.128& 0.467& 0.135& 0.294& 0.103& 0.348& 0.110& 0.309& 0.108 \\[2 mm]	
			\midrule
			(40, 20)& I& $\tau_{1}$& 1.047& 1047& 1.029& 1.051& 1.031& 1.012& 1.004& 0.986& 0.990& 0.949& 0.964 \\
			& & & 0.062& 0.062& 0.035& 0.063& 0.036& 0.052& 0.030& 0.053& 0.032& 0.052& 0.032   \\
			& & $\tau_{2}$& 1.571& 1.571& 1.543& 1.576& 1.549& 1.519& 1.507& 1.509& 1.492& 1.473& 1.484 \\
			& & & 0.139& 0.139& 0.078& 0.142& 0.080& 0.117& 0.075& 0.124& 0.076& 0.118& 0.076   \\[2 mm]
			& II& $\tau_{1}$& 1.016& 1.016& 1.008& 1.021& 1.060& 0.979& 0.984& 0.949& 0.962& 0.909& 0.928  \\
			& & & 0.087& 0.087& 0.056& 0.088& 0.057& 0.077& 0.051& 0.080& 0.052& 0.081& 0.053 \\
			& & $\tau_{2}$& 1.655& 1.655& 1.610& 1.661& 1.623& 1.594& 1.567& 1.587& 1.549& 1.547& 1.523 \\
			& & & 0.220& 0.220& 0.090& 0.225& 0.095& 0.182& 0.087& 0.189& 0.089& 0.176& 0.090 \\[2 mm]
			& III& $\tau_{1}$& 1.010& 1.010& 0.995& 1.013& 0.998& 0.972& 0.970& 0.941& 0.939& 0.902& 0.915  \\
			& & & 0.087& 0.087&  0.054& 0.088& 0.055& 0.077& 0.046& 0.081& 0.052& 0.083& 0.053  \\
			& & $\tau_{2}$& 1.650& 1.650& 1.583& 1.657& 1.598& 1.590& 1.527& 1.582& 1.525& 1.543& 1.491 \\
			& & & 0.223& 0.223& 0.103& 0.228& 0.109& 0.183& 0.086& 0.193& 0.080& 0.180& 0.079 \\
			\midrule	
			\end{tabular}}
			\end{center}
		\vspace{-0.5cm}
	\end{table}

\begin{table}[htbp!]
	\begin{center}
		\caption{ AL and CP of $95\%$ confidence and HPD credible intervals of $\tau_{1}$, and $\tau_{2}$ for different schemes with different values of $n$, $m$ when $T_{1}=0.5$, $T_{2}=1$, and $\Theta=(\tau_{1},\tau_{2})=(1,1.5)$. }
		\label{T4}
		\tabcolsep 7pt
		\small
			\begin{tabular}{*{10}c*{11}{r@{}l}}
				\toprule
				\multicolumn{3}{c}{} &
				\multicolumn{2}{c}{ACI} & \multicolumn{2}{c}{{Boot-$p$}} & \multicolumn{2}{c}{{Boot-$t$}} & \multicolumn{2}{c}{HPD} & \\
				\cmidrule(lr){4-5}\cmidrule(lr){6-7} \cmidrule(lr){8-9} \cmidrule(lr){10-11}
				\multicolumn{1}{c}{$(n,m)$}& \multicolumn{1}{c}{CS} & \multicolumn{1}{c}{$\Theta$}& \multicolumn{1}{c}{AL} & \multicolumn{1}{c}{CP}
				& \multicolumn{1}{c}{AL} & \multicolumn{1}{c}{CP}
				& \multicolumn{1}{c}{AL} & \multicolumn{1}{c}{CP} &  \multicolumn{1}{c}{AL} & \multicolumn{1}{c}{CP} \\	
				\midrule
                 (30, 10)& I& $\tau_{1}$ & 2.160& 0.992& {1.406}& {0.958}&
                 {1.515}& {0.955}& {0.745}& {0.969} \\[1 mm]
                 & & $\tau_{2}$ & 2.645& 0.980& {1.867}& {0.961}&
                 {1.922}& {0.957}& {1.352}& {0.976}\\[2 mm]
                 & II & $\tau_{1}$& 2.173& 0.995& {1.411}& {0.967}&
                 {1.543}& {0.962}& {0.700}& {0.978} \\[1 mm]
                 & & $\tau_{2}$& 2.662& 0.983& {2.201}& {0.963}&
                 {2.321}& {0.957}& {1.262}& {0.977} \\[2 mm]
                 & III & $\tau_{1}$& 2.198& 0.985& {1.394}& {0.958}& {1.428}& {0.953}&{0.756}& {0.966}\\[1 mm]
                 & & $\tau_{2}$& 2.747& 0.988& {2.374}& {0.954}&
                 {2.409}& {0.951}& {1.232}& {0.961} \\[2 mm]
                 \midrule
                  (30, 15)& I& $\tau_{1}$ & 1.705& 0.997& {1.099}& {0.962}& {1.205}& {0.960}& {0.695}& {0.975}\\[1 mm]
                  & & $\tau_{2}$ & 2.088& 0.989& {1.742}& {0.961}& {1.811}& {0.956}&{1.202}&{0.968}\\[2 mm]
                  & II& $\tau_{1}$ & 1.722& 0.939& {1.315}& {0.957}& {1.439}& {0.955}& {1.069}& {0.962} \\[1 mm]
                  & & $\tau_{2}$ & 2.222& 0.949& {2.043}& {0.955}& {2.124}& {0.952}& {1.711}& {0.960}  \\[2 mm]
                  & III& $\tau_{1}$ & 1.733& 0.940& {1.312}& {0.959}& {1.425}& {0.955}& {1.072}& {0.968} \\[1 mm]
                  & & $\tau_{2}$ & 2.235&  0.949& {2.025}& {0.957}& {2.025}&{0.954}&{1.781}& {0.965} \\[2 mm]
                  \midrule
                  (40, 10)& I& $\tau_{1}$ & 2.192& 0.992& {1.421}& {0.968}& {1.455}& {0.965}& {0.706}& {0.977} \\[1 mm]
                  & & $\tau_{2}$& 2.685& 0.988& {2.295}& {0.962}& {2.341}& {0.960}& {1.271}& {0.968}\\[2 mm]
                  & II& $\tau_{1}$ & 2.158& 0.996& {1.408}& {0.970}& {1.433}& {0.967}& {0.721}& {0.976}\\[1 mm]
                  & & $\tau_{2}$ & 2.644&  0.985& {2.306}& {0.962}& {2.399}&{0.955}&{1.303}& {0.968}\\[2 mm]
                  & III& $\tau_{1}$ & 2.217& 0.987& {1.430}& {0.966}& {1.502}& {0.964}& {0.742} & {0.974}\\[1 mm]
                  & & $\tau_{2}$ & 2.729& 0.990& {2.299}& {0.965}& {2.310}&{0.963}&{1.228}& {0.980}\\[2 mm]
                  \midrule
                  (40, 20)& I& $\tau_{1}$ & 1.451&  0.996& {0.933}& {0.957}& {0.980}& {0.955}& {0.678}& {0.975} \\[1 mm]
                  & & $\tau_{2}$ & 1.777& 0.988& {1.412}& {0.964}& {1.508}& {0.960}& {1.135}& {0.973}\\[2 mm]
                  & II& $\tau_{1}$ & 1.499& 0.958& {1.173}& {0.962}& {1.210}& {0.958}& {1.045}& {0.968}\\[1 mm]
                   & & $\tau_{2}$ & 1.913& 0.959& {1.572}& {0.961}&
                   {1.605}& {0.958}& {1.217}& {0.965}\\[2 mm]
                   & III & $\tau_{1}$ & 1.493&  0.955& {1.162}& {0.958}&
                   {1.198}& {0.954}& {1.063}& {0.970}\\[1 mm]
                   & & $\tau_{2}$ & 1.911& 0.957& {1.760}& {0.960}&
                   {1.804}& {0.954}& {1.476}& {0.966}\\[2 mm]
                   \midrule
	\end{tabular}
\end{center}
\vspace{-0.5cm}
\end{table}

\begin{table}[htbp!]
	\begin{center}
		\caption{Average values and MSEs of MLEs and Bayes estimates for different schemes with different values of $n$, $m$ when $T_{1}=1$, $T_{2}=1.5$, and $\Theta=(\tau_{1},\tau_{2})=(0.6,0.8)$. }
		\label{T5}
		\tabcolsep 7pt
		\small
		\scalebox{0.74}{
			\begin{tabular}{*{14}c*{13}{r@{}l}}
				\toprule
				\multicolumn{4}{c}{} &
				\multicolumn{2}{c}{SELF} & \multicolumn{2}{c}{LLF ($p=-0.05$)} & \multicolumn{2}{c}{LLF ($p=0.5$)}  &
				\multicolumn{2}{c}{GELF ($q=0.05$)} & \multicolumn{2}{c}{GELF ($q=0.5$)}  &\\
				\cmidrule(lr){5-6}\cmidrule(lr){7-8} \cmidrule(lr){9-10} \cmidrule(lr){11-12} \cmidrule(lr){13-14}
				\multicolumn{1}{c}{$(n,m)$}&\multicolumn{1}{c}{CS} & \multicolumn{1}{c}{$\Theta$}& \multicolumn{1}{c}{MLE} & \multicolumn{1}{c}{NIP}
				& \multicolumn{1}{c}{IP} & \multicolumn{1}{c}{NIP}
				& \multicolumn{1}{c}{IP} & \multicolumn{1}{c}{NIP}
				& \multicolumn{1}{c}{IP} & \multicolumn{1}{c}{NIP}
				& \multicolumn{1}{c}{IP} &\multicolumn{1}{c}{NIP}
				& \multicolumn{1}{c}{IP} \\
				\midrule
                 (30, 10)& I& $\tau_{1}$ & 0.624& 0.624& 0.596& 0.626& 0.598& 0.597& 0.583& 0.551& 0.542& 0.508& 0.534 \\
                 & & & 0.051& 0.051& 0.012& 0.052& 0.013& 0.042& 0.011& 0.042& 0.012& 0.042& 0.013 \\
                 & & $\tau_{2}$ & 0.935& 0.935& 0.856& 0.939& 0.860& 0.896& 0.833& 0.862& 0.825 & 0.819& 0.793  \\
                 & & & 0.132& 0.132& 0.028& 0.135& 0.030& 0.104& 0.025& 0.101& 0.024& 0.088& 0.022  \\[2 mm]
                 &  II & $\tau_{1}$ & 0.634& 0.634& 0.607& 0.636& 0.613& 0.607& 0.594& 0.560& 0.565& 0.516& 0.545 \\
                 & & & 0.045& 0.045& 0.021& 0.046& 0.021& 0.036& 0.020& 0.036& 0.021& 0.036& 0.022 \\
                 & & $\tau_{2}$ & 0.950& 0.950& 0.867& 0.954& 0.870& 0.911& 0.841& 0.876& 0.818& 0.832& 0.803  \\
                 & & & 0.121& 0.121& 0.047& 0.124& 0.049& 0.094& 0.045& 0.089& 0.043& 0.076& 0.041  \\  [2 mm]
                  &  III & $\tau_{1}$ & 0.609& 0.609& 0.589& 0.612& 0.591& 0.584& 0.576& 0.537& 0.546&  0.494& 0.525 \\
                  & & & 0.051& 0.051& 0.015&0.052& 0.016&  0.043& 0.014& 0.045& 0.015& 0.046& 0.016 \\
                  & & $\tau_{2}$ & 0.938& 0.938& 0.855& 0.943& 0.862& 0.899& 0.836& 0.865& 0.810& 0.821& 0.791 \\
                  & & &  0.122& 0.122& 0.029& 0.125& 0.030& 0.096& 0.026& 0.092& 0.024& 0.082& 0.023  \\[2 mm] 
                  \midrule
                  (30, 15)& I& $\tau_{1}$ & 0.597& 0.597& 0.586& 0.599& 0.587& 0.582& 0.586& 0.551& 0.559& 0.524& 0.542 \\
                  & & & 0.029& 0.029& 0.011& 0.030& 0.012& 0.026& 0.010& 0.027& 0.011& 0.028& 0.012 \\
                  & & $\tau_{2}$ & 0.896& 0.896& 0.842& 0.899& 0.849& 0.873& 0.834& 0.849& 0.816& 0.822& 0.804 \\
                  & & & 0.074& 0.074& 0.025& 0.075& 0.026& 0.063& 0.021& 0.060& 0.020& 0.055& 0.019 \\[2 mm]
                  & II & $\tau_{1}$ & 0.591& 0.591& 0.578& 0.592& 0.580& 0.574& 0.566& 0.543& 0.546& 0.515& 0.529\\
                  & & & 0.040& 0.040& 0.018& 0.042& 0.019& 0.037& 0.017& 0.038& 0.018& 0.039& 0.019  \\
                  & & $\tau_{2}$ & 0.857& 0.857& 0.816& 0.859& 0.818& 0.833& 0.802& 0.804& 0.780& 0.781& 0.767\\
                  & & & 0.090& 0.090& 0.040& 0.091& 0.041& 0.080& 0.035& 0.078& 0.034& 0.073& 0.033   \\[2 mm]
                 &  III & $\tau_{1}$ & 0.578& 0.578& 0.568& 0.581& 0.571& 0.562& 0.560& 0.532& 0.548& 0.504& 0.536  \\
                 & & & 0.048& 0.048& 0.014& 0.049& 0.015& 0.045& 0.014& 0.046& 0.015& 0.046& 0.015 \\
                 & & $\tau_{2}$ & 0.839& 0.839& 0.808& 0.841& 0.812& 0.815& 0.796& 0.792& 0.784& 0.764& 0.775  \\
                 & & & 0.103& 0.103& 0.025& 0.104& 0.026& 0.093& 0.022& 0.092& 0.023& 0.087& 0.022  \\[2 mm]
                 \midrule
                 (40, 10)& I& $\tau_{1}$ & 0.619& 0.619& 0.596& 0.621& 0.598& 0.594& 0.584& 0.547& 0.558& 0.505& 0.536 \\
                 & & & 0.047& 0.047& 0.012& 0.048& 0.013& 0.039& 0.012& 0.039& 0.012& 0.040&0.013  \\
                 & & $\tau_{2}$ & 0.929& 0.929& 0.853& 0.933& 0.856& 0.891& 0.834& 0.856& 0.811& 0.812& 0.795  \\
                 & & & 0.122& 0.122& 0.027& 0.125& 0.029& 0.096& 0.024& 0.093& 0.023& 0.081& 0.021 \\[2 mm]
                 & II & $\tau_{1}$ & 0.622& 0.622& 0.597& 0.624& 0.604& 0.596& 0.584& 0.550& 0.553& 0.507& 0.533 \\
                 & & & 0.045& 0.045& 0.012& 0.046& 0.013& 0.037& 0.010& 0.037& 0.011& 0.038& 0.012 \\
                 & & $\tau_{2}$ & 0.933& 0.933& 0.853& 0.937& 0.858& 0.894& 0.836& 0.861& 0.812& 0.817& 0.791 \\
                 & & & 0.118& 0.118& 0.025& 0.123& 0.026& 0.093& 0.023& 0.088& 0.022& 0.077& 0.021 \\[2 mm]
                &  III & $\tau_{1}$ & 0.620& 0.620& 0.593& 0.623& 0.596& 0.594& 0.585& 0.548& 0.557& 0.504& 0.536  \\
                & & & 0.049& 0.049& 0.012& 0.050& 0.013& 0.041& 0.011& 0.041& 0.011& 0.042& 0.012 \\
                & & $\tau_{2}$ & 0.937& 0.937& 0.852& 0.940& 0.856& 0.898& 0.834& 0.863& 0.810& 0.821& 0.789 \\
                & & & 0.124& 0.124& 0.026& 0.128& 0.029& 0.098& 0.023& 0.094& 0.022& 0.081& 0.021 \\[2 mm]
                \midrule
              (40, 20)& I& $\tau_{1}$ & 0.582& 0.582& 0.590& 0.584& 0.595& 0.572& 0.583& 0.548& 0.557& 0.528&  0.550\\
              & & & 0.017& 0.017& 0.008& 0.018& 0.009& 0.016& 0.008& 0.018& 0.009& 0.019& 0.009 \\
              & & $\tau_{2}$ & 0.874& 0.874& 0.848& 0.877& 0.851& 0.858& 0.841& 0.840& 0.827& 0.819& 0.812  \\
              & & & 0.043& 0.043& 0.022& 0.044& 0.023& 0.038& 0.020& 0.037& 0.019& 0.034& 0.018 \\[2 mm]
               & II & $\tau_{1}$ & 0.559& 0.559& 0.565& 0.562& 0.566& 0.548& 0.553& 0.524& 0.533& 0.505& 0.525 \\
               & & & 0.040& 0.040& 0.011& 0.041& 0.012& 0.034& 0.009& 0.036& 0.010& 0.037& 0.010 \\
               & & $\tau_{2}$ & 0.819& 0.819& 0.804& 0.821& 0.808& 0.803& 0.794& 0.784& 0.790& 0.764& 0.778 \\
               & & & 0.081& 0.081& 0.023& 0.082& 0.024& 0.079& 0.021& 0.077& 0.021& 0.073& 0.020 \\[2 mm]
               &  III & $\tau_{1}$ & 0.574& 0.574& 0.570& 0.575& 0.571& 0.562& 0.559& 0.539& 0.537& 0.518& 0.526\\
               & & & 0.039& 0.039& 0.010& 0.040& 0.011& 0.036& 0.008& 0.038& 0.009& 0.039& 0.010 \\
               & & $\tau_{2}$ & 0.841& 0.841& 0.810& 0.843& 0.812& 0.824& 0.803& 0.806& 0.785& 0.785& 0.772 \\
               & & & 0.082& 0.082& 0.022& 0.083& 0.023& 0.076& 0.020& 0.074& 0.019& 0.071& 0.018 \\[2 mm]
               \midrule
	\end{tabular}}
\end{center}
\vspace{-0.5cm}
\end{table}

\begin{table}[htbp!]
	\begin{center}
		\caption{ AL and CP of $95\%$ confidence and HPD credible intervals of $\tau_{1}$, and $\tau_{2}$ for different schemes with different values of $n$, $m$ when $T_{1}=1$, $T_{2}=1.5$, and $\Theta=(\tau_{1},\tau_{2})=(0.6,0.8)$. }
		\label{T6}
		\tabcolsep 7pt
		\small
			\begin{tabular}{*{10}c*{11}{r@{}l}}
				\toprule
				\multicolumn{3}{c}{} &
				\multicolumn{2}{c}{ACI} & \multicolumn{2}{c}{{Boot-$p$}} & \multicolumn{2}{c}{{Boot-$t$}} & \multicolumn{2}{c}{HPD} & \\
				\cmidrule(lr){4-5}\cmidrule(lr){6-7} \cmidrule(lr){8-9} \cmidrule(lr){10-11}
				\multicolumn{1}{c}{$(n,m)$}& \multicolumn{1}{c}{CS} & \multicolumn{1}{c}{$\Theta$}& \multicolumn{1}{c}{AL} & \multicolumn{1}{c}{CP}
				& \multicolumn{1}{c}{AL} & \multicolumn{1}{c}{CP}
				& \multicolumn{1}{c}{AL} & \multicolumn{1}{c}{CP} &  \multicolumn{1}{c}{AL} & \multicolumn{1}{c}{CP} \\
				\midrule
                (30, 10)& I& $\tau_{1}$ & 1.222& 0.981& {0.794}& {0.964}&
                {0.825}& {0.960}& {0.406}& {0.977} \\[1 mm]
                & &  $\tau_{2}$ & 1.497& 0.983& {1.291}& {0.962}&
                {1.330}& {0.959}& {1.176}& {0.965} \\[2 mm]
                & II & $\tau_{1}$ & 1.242& 0.991& {0.802}& {0.963}&
                {0.875}& {0.958}& {0.401}& {0.969}\\[1 mm]
                & & $\tau_{2}$ & 1.521& 0.994& {1.280}& {0.956}&
                {1.304}& {0.951}& {0.572}& {0.967}\\[2 mm]
                 & III & $\tau_{1}$ & 1.210& 0.966& {0.827}& {0.960}&
                 {0.878}& {0.957}& {0.502}& {0.977}\\[1 mm]
                 & & $\tau_{2}$ & 1.503& 0.981& {1.360}& {0.965}&
                 {1.386}& {0.961}& {0.652}& {0.978}\\[2 mm]
                 \midrule
                 (30, 15)& I& $\tau_{1}$ & 0.956& 0.986& {0.620}& {0.963}&
                 {0.695}& {0.958}& {0.410}& {0.976}\\[1 mm]
                 & & $\tau_{2}$ & 1.171& 0.991& {0.973}& {0.967}&
                 {0.985}& {0.965}& {0.592}& {0.978}\\[2 mm]
                 & II & $\tau_{1}$ & 0.960& 0.897& {0.815}& {0.945}&
                 {0.852}& {0.947}& {0.597}& {0.933} \\[1 mm]
                 & & $\tau_{2}$ & 1.156& 0.899& {1.296}& {0.957}&
                 {1.312}& {0.952}&{0.884}& {0.969} \\[2 mm]
                &  III & $\tau_{1}$ & 0.942& 0.870& {0.815}& {0.956}&
                {0.866}& {0.953}& {0.605}& {0.932}\\[1 mm]
                & & $\tau_{2}$ & 1.134& 0.876& {1.019}& {0.945}&
                {1.105}& {0.946}& {0.786}& {0.931} \\[2 mm]
                \midrule
                 (40, 10)& I& $\tau_{1}$ & 1.214& 0.982&{0.792}& {0.974}&
                 {0.824}& {0.969}& {0.428}&{0.980} \\[1 mm]
                 & & $\tau_{2}$ & 1.486& 0.984& {1.282}& {0.970}&
                 {1.297}& {0.965}& {0.611}& {0.981}\\[2 mm]
                  & II & $\tau_{1}$ & 1.219& 0.989& {0.788}& {0.965}& {0.806}& {0.961}& {0.424}& {0.978} \\[1 mm]
                  & & $\tau_{2}$ & 1.493& 0.994& {1.282}& {0.966}& {1.307}& {0.962}& {0.605}& {0.977}\\[2 mm]
                 &  III & $\tau_{1}$ & 1.219& 0.979& {0.806}& {0.964}&
                 {0.878}& {0.960}& {0.443}& {0.968}\\[1 mm]
                 & & $\tau_{2}$ & 1.499& 0.991& {1.293}& {0.965}&
                 {1.304}& {0.961}& {0.613}& {0.977}\\[2 mm]
                 \midrule
                 (40, 20)& I& $\tau_{1}$ & 0.807& 0.990& {0.521}& {0.970}& 
                 {0.578}& {0.965}&{0.406}& {0.981} \\[1 mm]
                  & & $\tau_{2}$ & 0.988& 0.997& {0.813}& {0.968}& {0.901}& {0.963}& {0.590}& {0.976}\\[2 mm]
                 & II & $\tau_{1}$ & 0.793& 0.873& {0.733}& {0.957}&
                 {0.812}& {0.955}& {0.603}& {0.963} \\[1 mm]
                 & & $\tau_{2}$ & 0.961& 0.881& {0.920}& {0.958}&
                 {0.975}& {0.953}& {0.855}& {0.940}\\[2 mm]
                 &  III & $\tau_{1}$ & 0.812& 0.887& {0.737}& {0.941}&
                 {0.784}& {0.945}& {0.609}& {0.968}\\[1 mm]
                 & & $\tau_{2}$ & 0.983& 0.893& {0.917}& {0.952}&
                 {0.972}& {0.948}& {0.881}& {0.937}\\[2 mm]
                 \midrule
	\end{tabular}
\end{center}
\vspace{-0.5cm}
\end{table}

\begin{table}[htbp!]
	\begin{center}
		\caption{Average values and MSEs of MLEs and Bayes estimates for different schemes with different values of $n$, $m$ when $T_{1}=1$, $T_{2}=1.5$, and $\Theta=(\tau_{1},\tau_{2})=(1,1.5)$. }
		\label{T7}
		\tabcolsep 7pt
		\small
		\scalebox{0.74}{
			\begin{tabular}{*{14}c*{13}{r@{}l}}
				\toprule
				\multicolumn{4}{c}{} &
				\multicolumn{2}{c}{SELF} & \multicolumn{2}{c}{LLF ($p=-0.05$)} & \multicolumn{2}{c}{LLF ($p=0.5$)}  &
				\multicolumn{2}{c}{GELF ($q=-0.05$)} & \multicolumn{2}{c}{GELF ($q=0.5$)} \\ 
				\cmidrule(lr){5-6}\cmidrule(lr){7-8} \cmidrule(lr){9-10} \cmidrule(lr){11-12} \cmidrule(lr){13-14}
				\multicolumn{1}{c}{$(n,m)$}&\multicolumn{1}{c}{CS} & \multicolumn{1}{c}{$\Theta$}& \multicolumn{1}{c}{MLE} & \multicolumn{1}{c}{NIP}
				& \multicolumn{1}{c}{IP} & \multicolumn{1}{c}{NIP}
				& \multicolumn{1}{c}{IP} & \multicolumn{1}{c}{NIP}
				& \multicolumn{1}{c}{IP} & \multicolumn{1}{c}{NIP}
				& \multicolumn{1}{c}{IP} &\multicolumn{1}{c}{NIP}
				& \multicolumn{1}{c}{IP} \\			
				\midrule
				 (30, 10)& I& $\tau_{1}$ & 1.120& 1.120& 1.051& 1.129& 1.058& 1.040& 1.009& 0.990& 0.972& 0.913& 0.923 \\
				 & & & 0.179& 0.179& 0.052& 0.188& 0.055& 0.120& 0.051& 0.129& 0.053& 0.117& 0.053  \\
				 & & $\tau_{2}$& 1.680& 1.680& 1.575& 1.693& 1.586& 1.560& 1.514& 1.549& 1.490& 1.472& 1.453 \\
				 & & & 0.403& 0.403& 0.119& 0.422& 0.123& 0.272& 0.094& 0.317& 0.098& 0.285& 0.097  \\[2 mm]
				 & II& $\tau_{1}$ & 1.126& 1.126& 1.052& 1.135& 1.089& 1.046& 1.005& 0.985& 0.969& 0.917& 0.932 \\
				 & & & 0.174& 0.174& 0.051& 0.182& 0.052& 0.117& 0.042& 0.123& 0.044& 0.117& 0.045  \\
				 & & $\tau_{2}$& 1.689& 1.689& 1.580& 1.702& 1.598& 1.569& 1.513& 1.557& 1.496& 1.479& 1.449 \\
				 & & & 0.391& 0.391& 0.118& 0.409& 0.120& 0.264& 0.093& 0.305& 0.099& 0.273& 0.096  \\[2 mm]
				  & III& $\tau_{1}$ & 1.118& 1.118& 1.046& 1.127& 1.065& 1.039& 0.998& 0.988& 0.971& 0.910& 0.919 \\
				  & & & 0.175& 0.175& 0.055& 0.183& 0.058& 0.117& 0.050& 0.126& 0.052& 0.115& 0.050  \\
				  & & $\tau_{2}$&  1.676& 1.676& 1.568& 1.689& 1.581& 1.558& 1.509& 1.545& 1.486& 1.468& 1.464 \\
				  & & & 0.393& 0.393& 0.113& 0.412& 0.116& 0.265& 0.097& 0.310& 0.101& 0.279& 0.097 \\[2 mm]
				  \midrule
				  (30, 15)& I& $\tau_{1}$ & 1.086& 1.086& 1.045& 1.091& 1.053& 1.037& 1.011& 1.002& 0.997& 0.951& 0.958  \\
				  & & & 0.092& 0.092& 0.042& 0.094& 0.043& 0.071& 0.035& 0.072& 0.036& 0.067& 0.033 \\
				  & & $\tau_{2}$ & 1.629& 1.629& 1.566& 1.637& 1.580& 1.555& 1.512& 1.544& 1.497& 1.494& 1.470  \\
				  & & & 0.206& 0.206& 0.085& 0.212& 0.089& 0.160& 0.080& 0.172& 0.083& 0.159& 0.080   \\[2 mm]
				  & II & $\tau_{1}$ & 1.060& 1.060& 1.033& 1.066& 1.037& 1.013& 1.006& 0.976& 0.965& 0.927& 0.937 \\
				  & & & 0.093& 0.093& 0.045& 0.096& 0.047& 0.075& 0.040& 0.077& 0.041& 0.075& 0.041   \\
				  & & $\tau_{2}$ & 1.612& 1.612& 1.572& 1.620& 1.615& 1.540& 1.549& 1.528& 1.530& 1.478& 1.505  \\
				  & & & 0.205& 0.205& 0.091& 0.211& 0.093& 0.161& 0.084& 0.174& 0.086& 0.163& 0.084 \\[2 mm]
				   & III & $\tau_{1}$ & 1.059& 1.059& 1.028& 1.065& 1.033& 1.012& 1.003& 0.976& 0.981& 0.926& 0.937 \\
				   & & & 0.097& 0.097& 0.047& 0.099& 0.048& 0.078& 0.040& 0.080& 0.041& 0.078& 0.042  \\
				   & & $\tau_{2}$ & 1.609& 1.609& 1.560& 1.617& 1.563& 1.537& 1.518& 1.525& 1.498& 1.475& 1.466 \\
				   & & & 0.212& 0.212& 0.101& 0.217& 0.104& 0.167& 0.082& 0.180& 0.086& 0.164& 0.085 \\[2 mm]
				   \midrule
				   (40, 10)& I& $\tau_{1}$ & 1.120& 1.120& 1.052& 1.129& 1.060& 1.041& 1.008& 0.990& 0.955& 0.912& 0.916  \\
				   & & & 0.166& 0.166& 0.049& 0.174& 0.051& 0.112& 0.042& 0.119& 0.043& 0.108& 0.042 \\
				   & &  $\tau_{2}$ & 1.680& 1.680& 1.560& 1.693& 1.582& 1.562& 1.498& 1.549& 1.490& 1.472& 1.439 \\
				   & & & 0.374& 0.374& 0.112& 0.392& 0.116& 0.253& 0.090& 0.293& 0.096& 0.263& 0.092 \\[2 mm]
				   & II & $\tau_{1}$ & 1.117& 1.117& 1.043& 1.125& 1.049& 1.037& 0.995& 0.987& 0.964& 0.910& 0.926 \\
				   & & & 0.186& 0.186& 0.057& 0.195& 0.059& 0.125& 0.048& 0.135& 0.050& 0.123& 0.048 \\
				   & & $\tau_{2}$ & 1.674& 1.674& 1.560& 1.688& 1.565& 1.555& 1.490& 1.544& 1.468& 1.467& 1.435  \\
				   & & & 0.420& 0.420& 0.117& 0.439& 0.124& 0.282& 0.093& 0.333& 0.097& 0.299& 0.094 \\[2 mm]
				   & III & $\tau_{1}$ & 1.088& 1.088& 1.035& 1.096& 1.039& 1.014& 0.998& 0.962& 0.956& 0.887& 0.910   \\
				   & & & 0.146& 0.146& 0.048& 0.152& 0.050& 0.103& 0.042& 0.109& 0.043& 0.105& 0.043 \\
				   & & $\tau_{2}$& 1.633& 1.633& 1.551& 1.645& 1.560& 1.521& 1.494& 1.506& 1.473& 1.430& 1.423   \\
				   & & & 0.328& 0.328& 0.108& 0.342& 0.111& 0.230& 0.092& 0.263& 0.094& 0.243& 0.093  \\[2 mm]
				   \midrule
				   (40, 20)& I& $\tau_{1}$ & 1.052& 1.052& 1.029& 1.056& 1.032& 1.018& 1.004& 0.990& 0.983& 0.955& 0.956  \\
				   & & & 0.064& 0.064& 0.036& 0.066& 0.037& 0.053& 0.031& 0.054& 0.032& 0.052& 0.032  \\
				   & & $\tau_{2}$ & 1.578& 1.578& 1.542& 1.584& 1.545& 1.527& 1.504& 1.517& 1.493& 1.481& 1.469  \\
				   & & & 0.143& 0.143& 0.080& 0.146& 0.082& 0.119& 0.076& 0.127& 0.078& 0.121& 0.078  \\[2 mm]
				   & II & $\tau_{1}$ & 1.050& 1.050& 1.028& 1.054& 1.030& 1.015& 1.008& 0.988& 0.974& 0.951&  0.950 \\
				   & & & 0.073& 0.073& 0.043& 0.074& 0.044& 0.061& 0.039& 0.063& 0.040& 0.060& 0.039\\
				   & & $\tau_{2}$ & 1.595& 1.595& 1.554& 1.601& 1.559& 1.541& 1.513& 1.532& 1.507& 1.496& 1.489 \\
				   & & & 0.160& 0.160& 0.092& 0.163& 0.094& 0.133& 0.084& 0.140& 0.086& 0.133& 0.084 \\[2 mm]
				    & III & $\tau_{1}$ & 1.041& 1.041& 1.017& 1.045& 1.021& 1.007& 0.993& 0.979& 0.971& 0.943& 0.948  \\
				    & & &0.065& 0.065& 0.040& 0.066& 0.041& 0.055& 0.037& 0.057& 0.038& 0.056& 0.037\\
				    & &  $\tau_{2}$ & 1.583& 1.583& 1.545& 1.589& 1.550& 1.531& 1.526& 1.522& 1.511& 1.485& 1.498 \\
				    & & & 0.142& 0.142& 0.082 & 0.145& 0.085& 0.119& 0.076& 0.125& 0.078& 0.119& 0.077 \\[2 mm]
				    \midrule
		\end{tabular}}
\end{center}
\vspace{-0.5cm}
\end{table}

\begin{table}[htbp!]
	\begin{center}
		\caption{ AL and CP of $95\%$ confidence and HPD credible intervals of $\tau_{1}$, and $\tau_{2}$ for different schemes with different values of $n$, $m$ when $T_{1}=1$, $T_{2}=1.5$, and $\Theta=(\tau_{1},\tau_{2})=(1,1.5)$. }
		\label{T8}
		\tabcolsep 7pt
		\small
			\begin{tabular}{*{10}c*{11}{r@{}l}}
				\toprule
				\multicolumn{3}{c}{} &
				\multicolumn{2}{c}{ACI} & \multicolumn{2}{c}{{Boot-$p$}} & \multicolumn{2}{c}{{Boot-$t$}} & \multicolumn{2}{c}{HPD} & \\
				\cmidrule(lr){4-5}\cmidrule(lr){6-7} \cmidrule(lr){8-9} \cmidrule(lr){10-11}
				\multicolumn{1}{c}{$(n,m)$}&  \multicolumn{1}{c}{CS} & \multicolumn{1}{c}{$\Theta$}& \multicolumn{1}{c}{AL} & \multicolumn{1}{c}{CP}
				& \multicolumn{1}{c}{AL} & \multicolumn{1}{c}{CP}
				& \multicolumn{1}{c}{AL} & \multicolumn{1}{c}{CP} &  \multicolumn{1}{c}{AL} & \multicolumn{1}{c}{CP} \\
				\midrule
				(30, 10)& I& $\tau_{1}$ & 2.195& 0.993& {1.420}& {0.975}&
				{1.515}& {0.970}& {0.707}& {0.985}\\[1 mm]
				& & $\tau_{2}$& 2.688& 0.987& {2.306}& {0.968}&
				{2.415}& {0.964}& {1.269}& {0.977} \\[2 mm]
				& II& $\tau_{1}$ & 2.207& 0.996& {1.416}& {0.972}&
				{1.512}& {0.968}& {0.635}& {0.979}\\[1 mm]
				& & $\tau_{2}$ & 2.703& 0.984& {2.296}& {0.965}& 
				{2.335}& {0.962}&{1.165}& {0.972}\\[2 mm]
                 & III& $\tau_{1}$ & 2.190& 0.997& {1.410}& {0.971}&
                 {1.525}& {0.967}& {0.663}& {0.980}\\[1 mm]
                 & & $\tau_{2}$ & 2.683& 0.988& {2.289}& {0.971}&
                 {2.320}& {0.968}& {1.179}& {0.978} \\[2 mm]
                 \midrule
                 (30, 15)& I& $\tau_{1}$ & 1.738& 0.995& {1.101}& {0.968}&
                 {1.211}& {0.962}& {0.692}& {0.977}\\[1 mm]
                 & & $\tau_{1}$ & 2.129& 0.985& {1.743}& {0.969}& 
                 {1.818}& {0.966}&{1.194}& {0.978}\\[2 mm]
                 & II& $\tau_{1}$ & 1.708& 0.989& {1.108}& {0.968}&
                 {1.212}& {0.965}& {0.694}& {0.974}\\[1 mm]
                 & & $\tau_{2}$ & 2.106& 0.981& {1.742}& {0.974}&
                 {1.825}& {0.970}& {1.134}& {0.981}\\[2 mm]
                 & III& $\tau_{1}$ & 1.705& 0.989& {1.101}& {0.967}&
                 {1.225}& {0.964}& {0.701}& {0.974}\\[1 mm]
                 & & $\tau_{2}$ & 2.102& 0.985& {1.745}& {0.968}&
                 {1.810}& {0.965}& {1.130}& {0.973}\\[2 mm]
                 \midrule
                 (40, 10)& I& $\tau_{1}$ & 2.195& 0.995& {1.414}& {0.966}&
                 {1.551}& {0.961}& {0.744}& {0.972}\\[1 mm]
                 & & $\tau_{2}$ & 2.689& 0.984& {2.304}& {0.968}&
                 {2.415}& {0.965}& {1.337}& {0.976}\\[2 mm]
                 & II & $\tau_{1}$ & 2.188& 0.995& {1.419}& {0.966}& {1.524}& {0.962}&{0.678}& {0.976} \\[1 mm]
                 & & $\tau_{2}$ & 2.680& 0.985&  {2.285}& {0.961}& {2.312}& {0.959}& {1.218}& {0.968} \\[2 mm]
                 & III & $\tau_{1}$ & 2.133& 0.994& {1.417}& {0.968}&
                 {1.536}& {0.963}& {0.668}& {0.977} \\[1 mm]
                 & &  $\tau_{2}$ & 2.613& 0.983& {2.301}& {0.965}&
                 {2.454}& {0.960}& {1.195}& {0.970} \\[2 mm]
                 \midrule
                 (40, 20)& I& $\tau_{1}$ & 1.459& 0.993& {0.936}& {0.968}&
                 {0.965}& {0.966}& {0.647}& {0.979}\\[1 mm]
                 & &  $\tau_{2}$ & 1.786& 0.987& {1.449}& {0.965}&
                 {1.545}& {0.961}& {1.085}& {0.976}\\[2 mm]
                 & II& $\tau_{1}$ & 1.465& 0.987& {0.945}& {0.964}&
                 {0.994}& {0.959}& {0.668}& {0.975} \\[1 mm]
                 & & $\tau_{2}$ & 1.806& 0.985& {1.469}& {0.968}&
                 {1.515}& {0.963}& {1.068}& {0.974}\\[2 mm]
                 & III& $\tau_{1}$ & 1.453& 0.988& {0.949}& {0.957}&
                 {1.025}& {0.955}& {0.633}& {0.963} \\[1 mm]
                 & & $\tau_{2}$ & 1.793& 0.987& {1.466}& {0.964}&
                 {1.526}& {0.961}& {1.024}& {0.972} \\[2 mm]
                 \midrule
	\end{tabular}
\end{center}
\vspace{-0.5cm}
\end{table}	

\begin{figure}[h!]
	\begin{center}
		\subfigure[]{\label{c1}\includegraphics[height=2 in, width = 3 in]{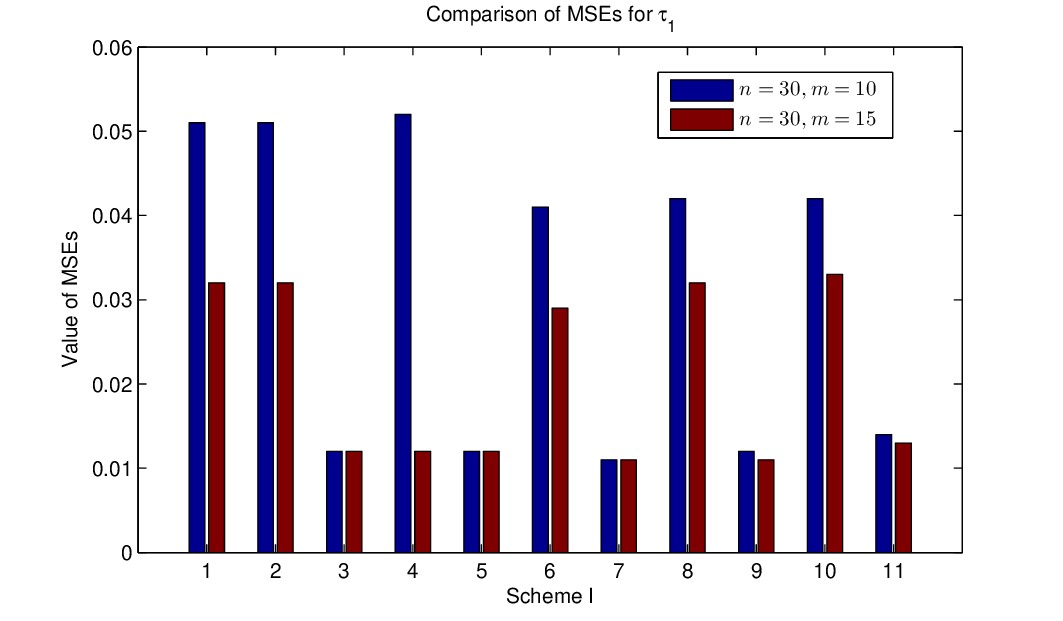}}
		\subfigure[]{\label{c1}\includegraphics[height=2 in, width= 3 in]{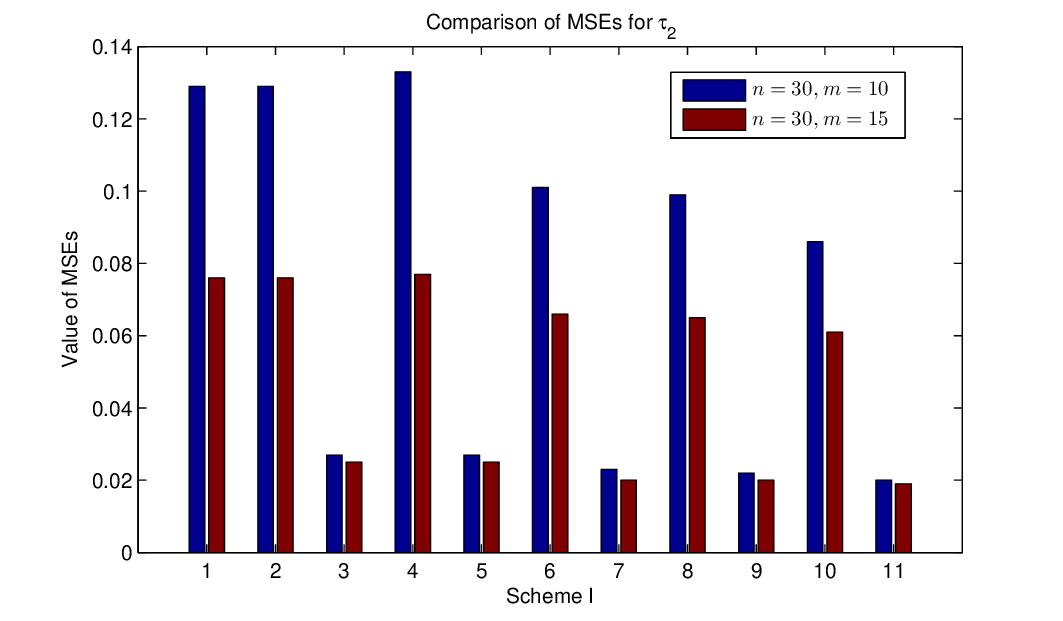}}
		\subfigure[]{\label{c1}\includegraphics[height=2 in, width = 3 in]{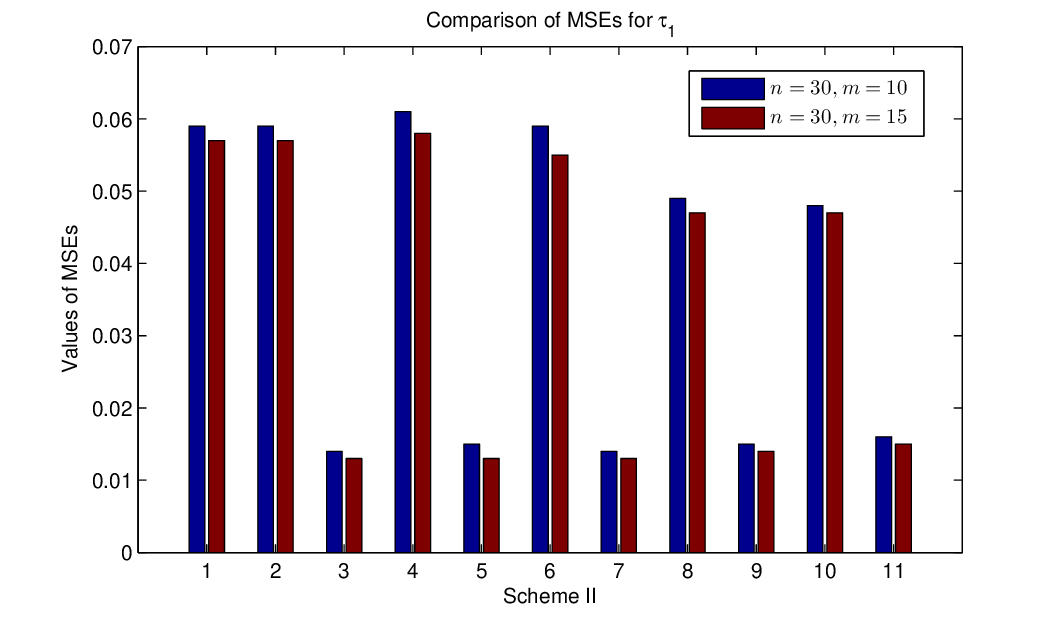}}
		\subfigure[]{\label{c1}\includegraphics[height=2 in, width= 3 in]{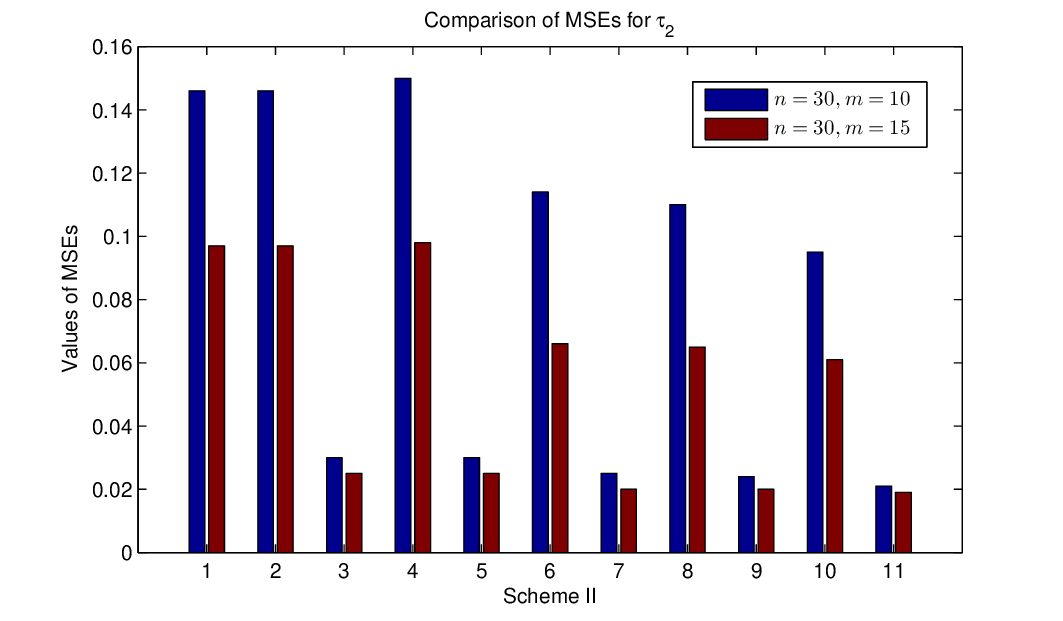}}
		\subfigure[]{\label{c1}\includegraphics[height=2 in, width = 3 in]{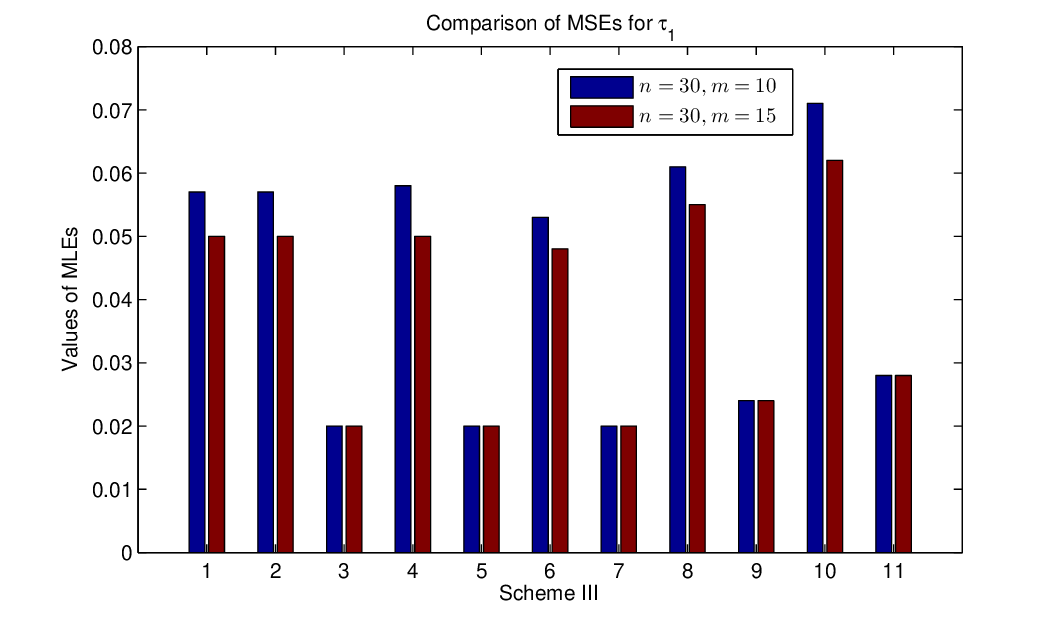}}
		\subfigure[]{\label{c1}\includegraphics[height=2 in, width= 3 in]{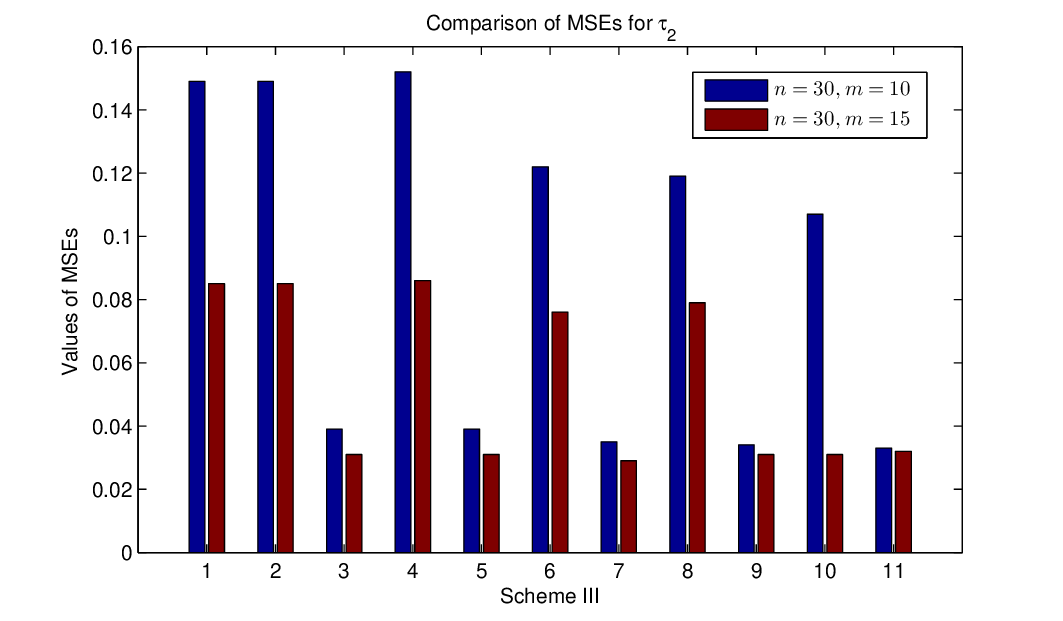}}
		\caption{ Bar plots of MSEs of the proposed estimates when $T_1=0.5$, $T_2=1$, $\tau_{1}=0.6$ and $\tau_{2}=0.8$.  }
	\end{center}
\end{figure}

\begin{table}[htbp!]
\begin{center}	
	\caption{ Breaking strengths of jute fiber data.}
	\label{T9}
	\tabcolsep 7pt
	\small
	\begin{tabular}{*{10}c*{11}{r@{}l}}
	\toprule
	\multicolumn{1}{c}{Cause of death}&   \multicolumn{1}{c}{Data} & \\
	\midrule
	\mbox{Cause 1}& {43.93, 50.16, 101.15, 123.06, 108.94, 141.38, 151.48, 163.40, 177.25, 183.16,~~~~}\\
	&	{212.13, 257.44, 262.90, 291.27, 303.90, 323.83, 353.24, 376.42, 383.43, 422.11,}\\
	& {506.60, 530.55, 590.48, 637.66, 671.49, 693.73, 700.74, 704.66, 727.23, 778.17.}\\ [0.2 cm]
	\mbox{Cause 2}& {36.75, 45.58, 48.01, 71.46, 83.55, 99.72, 113.85, 116.99, 119.86, 145.96, 166.49}\\
	& {187.13, 187.85, 200.16, 244.53, 284.64, 350.70, 375.81, 419.02, 456.60, 547.44,} \\
	& {578.62, 581.60, 585.57, 594.29, 662.66, 688.16, 707.36, 756.70, 765.14.~~~~~~~~~~} \\
	\bottomrule
		\end{tabular}
\end{center}
\vspace{-0.5cm}
\end{table}	
	
\section{Optimality criteria}
In reliability and survival analysis, an optimum censoring plan among chosen schemes is desired to get a sufficient amount of information about the unknown model parameters. Here, three commonly used criteria have been considered based on the variance-covariance matrix (VCM) of the observed Fisher information matrix corresponding to the MLEs of unknown parameters (See Table $10$). In the statistical literature, $A$ and $D$-optimality criteria have been widely used. Here, these two criteria intend to minimize the trace and determinant of VCM, respectively. Further, $F$-optimality intends to minimize the trace of the observed Fisher information matrix of the MLEs. According to these optimality criteria, the corresponding optimal censoring plans have been considered in Table $12$.
\begin{table}[htbp!]
	\begin{center}
		\caption{Different optimality criterion.}
		\label{T10}
		\tabcolsep 7pt
		\small
		\scalebox{0.8}{
			\begin{tabular}{*{3}c*{2}{r@{}l}}
				\toprule
				\multicolumn{1}{c}{Criterion}&&& & &\multicolumn{1}{c}{Goal}  \\
				\midrule
				A-optimality &&& & & minimum trace $(I^{-1}(\widehat{\Theta}))$&\\
				D-optimality &&& & &minimum det $(I^{-1}(\widehat{\Theta}))$~~~& \\
				F-optimality &&&&& maximum  trace $(I(\widehat{\Theta}))$~~~&\\				
				\bottomrule
		\end{tabular}}
	\end{center}
	\vspace{-0.5cm}
\end{table}

\section{Real data analysis}

In this section, in order to show the applicability of the proposed methods, a real life data set has been analyzed. The data set (reported by Xia et. al \cite{xia2009study}) is presented in Table $\ref{T9}$, contains breaking strengths of jute fiber in which failure caused by two different gauge lengths $10$ mm and $20$ mm. First, we have to check whether these data set can be used for the exponential distribution or not. By using Kolmogorov–Smirnov (KS) test for these two data sets, the KS distance and corresponding $p$-values (within bracket) are obtained as $0.1723$ $(0.2995)$ and $0.1289$ $(0.6541)$. From this result, we can conclude that exponential distribution can be chosen as a reasonable model for these two data sets.
	\begin{figure}[h!]
		\begin{center}
			\subfigure[]{\label{c1}\includegraphics[ width = 2 in]{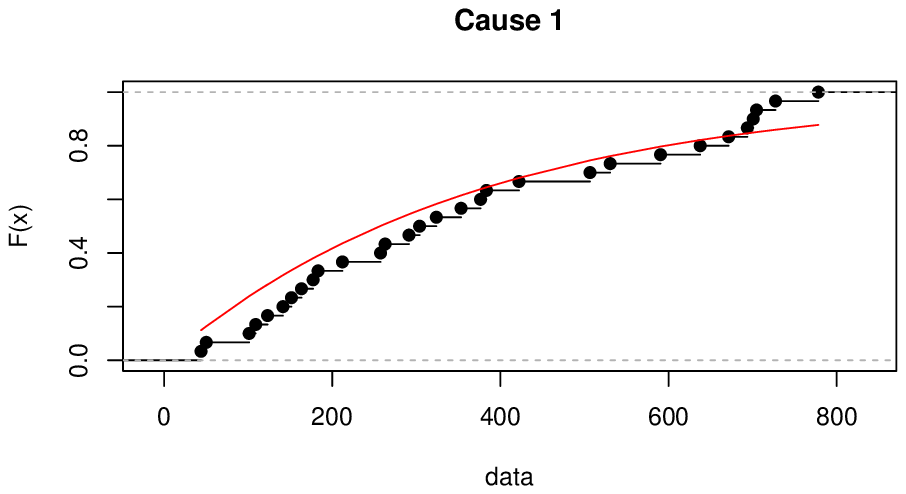}}
			\subfigure[]{\label{c1}\includegraphics[width= 2 in]{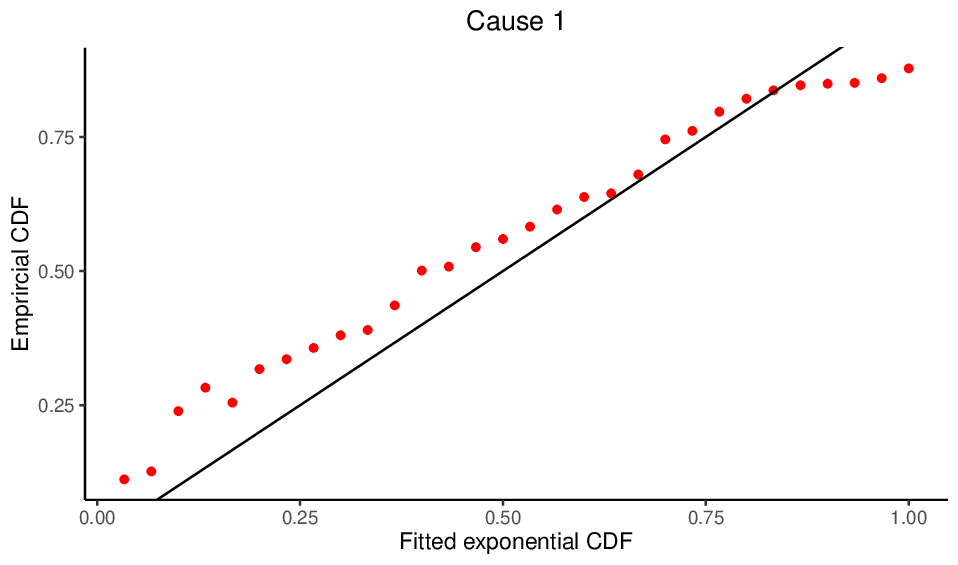}}
			\subfigure[]{\label{c1}\includegraphics[width= 2 in]{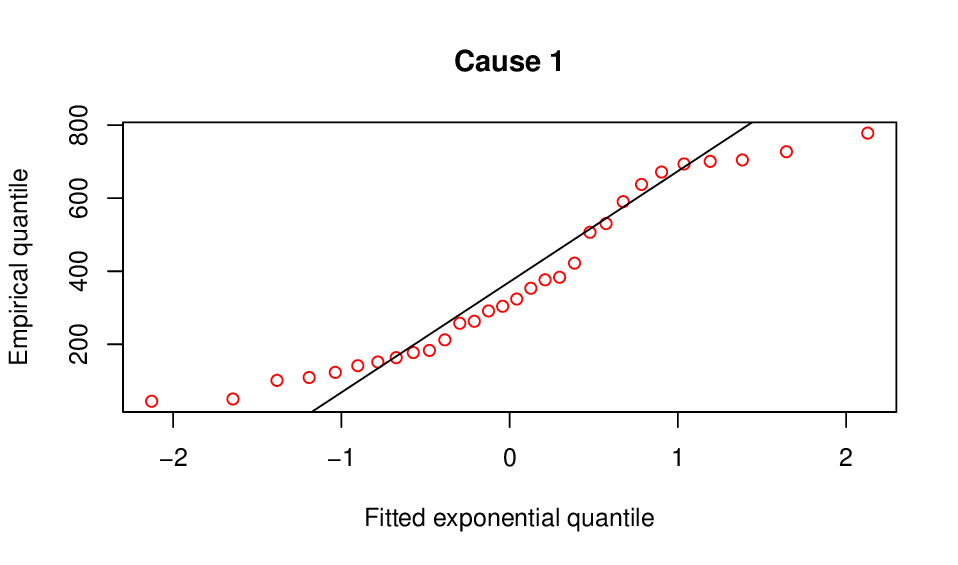}}
			\subfigure[]{\label{c1}\includegraphics[ width = 2 in]{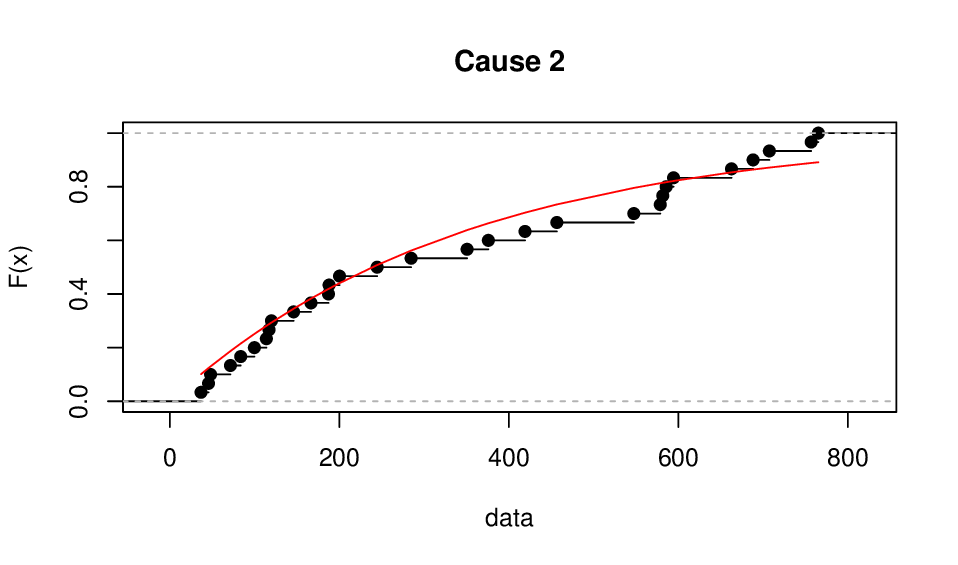}}
			\subfigure[]{\label{c1}\includegraphics[width= 2 in]{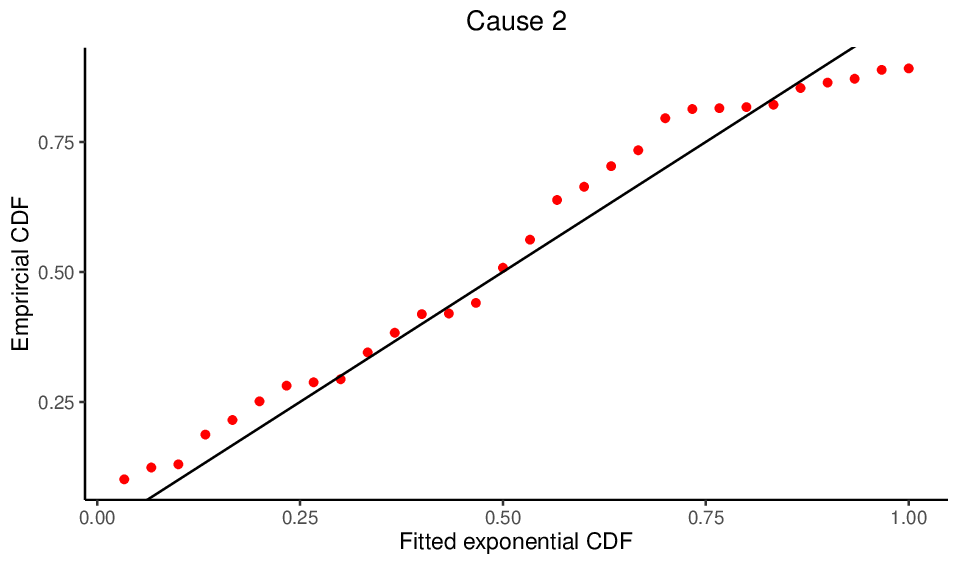}}
			\subfigure[]{\label{c1}\includegraphics[width= 2 in]{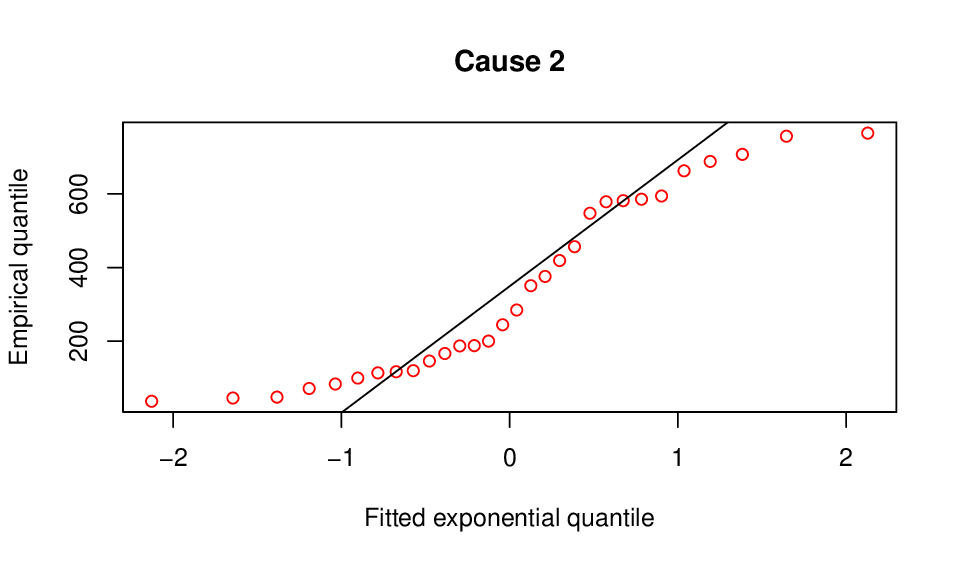}}
			\caption{ ECDF, P-P and Q-Q plots for breaking strength of jute fiber data. }
		\end{center}
	\end{figure}
	 Further, empirical cumulative distribution function (ECDF) plots, probability-probability (P-P) plots and quantile-quantile (Q-Q) plots are presented in Figure $3$ which support that exponential distribution is reasonable to fit these data sets.
   Using different censoring schemes, we have generated two  improved adaptive type-II progressively censored competing risks failure data sets from Table $\ref{T9}$ with $n=60$, $m=30$, $T_1=200$, and $T_2=350$. This censored competing risks data are tabulated in Table $11$.\\
\begin{table}
	\begin{center}
		\caption{ Improved adaptive type-II progressively censored competing risks data set with two causes of failues.}
		\label{T11}
		\tabcolsep 7pt
		\small
		\scalebox{0.9}{
		\begin{tabular}{*{1}c*{1}{r@{}l}}
	    \toprule
	    \textbf{Data I:}  $T=291.27$ and $D=30$\\
	    \midrule
	    (36.75,2), (43.93,1), (45.58,2), (48.01,2), (50.16,1), (71.46,2), (83.55,2), (99.72,2), (101.15,1),(108.94,1),\\ 
	     (113.85,2), (116.99,2),(119.86,2), (123.06,1), (141.38,1), (145.96,2), (151.48,1), (163.40,1), (166.49,2),\\
	      (177.25,1), (183.16,1),
	      (187.13,2), (187.85,2), (200.16,2), (212.13,1), (244.53,2), (257.44,1), (262.90,1),\\
	       (284.64,2), (291.27,1).~~~~~~~~~~~~~~~~~~~~~~~~~~~~~~~~~~~~~~~~~~~~~~~~~~~~~~~~~~~~~~~~~~~~~~~~~~~~~~~~~~~~~~~~~~~~~~~~~~~~~~~~~\\
	    \midrule
	    \textbf{Data II:} $T=350$ and $D=27$\\
	    \midrule
	    (43.93,1), (45.58,2), (50.16,1), (71.46,2), (83.55,2), (99.72,2), (101.15,1), (108.94,1), (113.85,2),(119.86,2),\\
	       (123.06,1), (141.38,1), (145.96,2),  (163.40,1), (166.49,2), (177.25,1), (187.13,2), (187.85,2), (200.16,2),~~~~\\
	      (212.13,1), (244.53,2), (257.44,1), (262.90,1), (284.64,2), (291.27,1), (303.90,1), (323.83,1).~~~~~~~~~~~~~~~~~~~~\\
	     \bottomrule
\end{tabular}}
\end{center}
\vspace{-0.5cm}
\end{table}

\begin{table}[htbp!]
	\begin{center}
			\caption{ Point and interval estimates based on real data set.}
			\label{T12}
			\tabcolsep 7pt
			\small
			\scalebox{0.73}{
			\begin{tabular}{*{10}c*{11}{r@{}l}}
				\toprule
				\multicolumn{1}{c}{Data} & \multicolumn{1}{c}{CS} & \multicolumn{1}{c}{Estimates} &
				\multicolumn{1}{c}{$\tau_{1}$} & \multicolumn{1}{c}{$\tau_{2}$}&  \multicolumn{1}{c}{A-optimality}& \multicolumn{1}{c}{D-optimality}&
				\multicolumn{1}{c}{F-optimality} \\
				\midrule
				I& 	I& MLE& 0.00147& 0.00161& \textbf{3.653$\times 10^{-8}$}& \textbf{3.336$\times 10^{-16}$}& \textbf{1.094$\times 10^{8}$}\\
				& & Bayes& 0.00197& 0.00200 &\\
				& & ACI& (0.00120, 0.00173)[0.00053]& (0.00135, 0.00188)[0.00053] &\\
				& & Boot-$p$& (0.00197, 0.00404)[0.00207]& (0.00215, 0.00463)[0.00248] &\\
				& & Boot-$t$& (0.00198, 0.00414)[0.00216]& (0.00223, 0.00476)[0.00253] &\\
				& & HPD& (0.00177, 0.00221)[0.00044]& (0.00179, 0.00224)[0.00045] &\\ 
				\midrule
				& 	II& MLE& 0.00204& 0.00228& 2.642$\times 10^{-7}$& 1.585$\times 10^{-14}$& 1.666$\times 10^{7}$\\
				& & Bayes& 0.00293& 0.00305 &\\
				& & ACI& (0.00145, 0.00264)[0.00119]& (0.00147, 0.00309)[0.00162] &\\
				& & Boot-$p$& (0.00132, 0.00312)[0.00180]& (0.00159, 0.00371)[0.00212] &\\
				& & Boot-$t$& (0.00133, 0.00324)[0.00191]& (0.00143, 0.00382)[0.00239] &\\
				& & HPD& (0.00234, 0.00342)[0.00108]& (0.00256, 0.00355)[0.00099] &\\
				\midrule
				II& I& MLE& 0.00125& 0.00129& \textbf{3.992$\times 10^{-8}$}& \textbf{3.984$\times 10^{-16}$}& \textbf{1.002$\times 10^{8}$}\\
				& & Bayes& 0.00140& 0.00142 &\\
				& & ACI& (0.00097, 0.00152)[0.00055]& (0.00101, 0.00157)[0.00056] &\\
				& & Boot-$p$& (0.00078, 0.00192)[0.00114]& (0.00112, 0.00214)[0.00102] &\\
				& & Boot-$t$& (0.00082, 0.00204)[0.00122]& (0.00081, 0.00215)[0.00134] &\\
				& & HPD& (0.00134, 0.00176)[0.00042]& (0.00139, 0.00170)[0.00031] &\\
				\midrule
				& II& MLE& 0.00157& 0.00166& 4.078$\times 10^{-8}$& 4.158$\times 10^{-16}$& 0.981$\times 10^{8}$\\
				& & Bayes& 0.00224& 0.00229 &\\
				& & ACI& (0.00129, 0.00186)[0.00057]& (0.00138, 0.00194)[0.00056] &\\
				& & Boot-$p$& (0.00106, 0.00198)[0.00092]& (0.00122, 0.00215)[0.00093] &\\
				& & Boot-$t$& (0.00112, 0.00213)[0.00102]& (0.00113, 0.00213)[0.00100] &\\
				& & HPD& (0.00208, 0.00256)[0.00048]& (0.00209, 0.00259)[0.00050] &\\
				\bottomrule
			\end{tabular}}
		\end{center}
	\vspace{-0.5cm}
\end{table}	
Based on the competing risks data shown in Table $11$, point and interval estimates have been computed and tabulated in Table $12$. In case of interval estimates, the significance level ($\gamma$) has been considered as 0.05 and the interval lengths are provided in square brackets. To compute the Bayes estimates, NIP has been adopted under SELF. From Table $12$, it has been observed that the point estimates based on MLEs and Bayes estimates are pretty close to each other. The HPD credible intervals perform better than ACIs and bootstrap intervals. We have also computed three different optimality criteria for these two data sets based on censoring scheme (CS) I and II. Based on these criteria, it has been observed that CS-I is optimal than CS-II for both of these two data sets.

\section{Conclusion}
 \noindent In this paper, statistical inferences have been proposed based on classical and Bayesian approaches under IAT-II PCS in the presence of competing risks data. Here, we have considered two causes of failures. It is assumed that the latent failure times due to different causes are exponentially distributed with different scale parameters. Bayes estimates are obtained under both non-informative and informative priors. We have constructed confidence intervals using the asymptotic normality property of MLEs for the unknown parameters. Also, two different parametric bootstrap confidence intervals have been constructed. The HPD credible intervals are obtained from the posterior density functions. To obtain the optimal censoring plan, three different optimality criteria such as $A$, $D$ and $F$-optimality have been computed. In simulation studies, it is noticed that the Bayes estimates perform better than MLEs. The Bayes estimates with respect to IP perform better than that with respect to NIP. The HPD credible intervals perform better than asymptotic and bootstrap confidence intervals in terms of AL. In terms of CP, boot-$t$ confidence intervals perform better than other interval estimates. A real-life data set is also considered for further illustrations. \\\\
 \textbf{Acknowledgement:} The authors would like to thank the Editor in Chief, an Associate
 Editor and two anonymous reviewers for their positive remarks and useful comments.
 The author S. Dutta, thanks the Council of Scientific and Industrial Research (C.S.I.R.
 Grant No. 09/983(0038)/2019-EMR-I), India, for the financial assistantship received to carry out this
 research work. Both the authors thanks the research facilities received from the Department of Mathematics, National Institute of Technology Rourkela, India.\\
\bibliography{myref1}
\end{document}